\documentclass[sigconf,nonacm]{acmart}

\usepackage{url}            % simple URL typesetting

\usepackage{booktabs}       % professional-quality tables
\usepackage{amsfonts}       % blackboard math symbols
\usepackage{nicefrac}       % compact symbols for 1/2, etc.
\usepackage{microtype}      % microtypography
\usepackage{lipsum}

\usepackage{listings}  % code block
\usepackage{mdframed}

\usepackage{titlesec}
\titleformat{\section}
  {\normalfont\Large\bfseries}{\thesection}{1em}{\MakeUppercase}
\titlespacing{\subsubsection}
  {0pt}      % 左边缩进
  {2pt}     % 上方间距（你要调整的值）
  {0pt}      % 下方间距（标题和正文之间的距离）
  
\usepackage{amsmath,bm}     % bold
\usepackage{enumerate}
\usepackage{pifont}     % more styles for bullets
\usepackage{enumitem}
\usepackage{xcolor}

\usepackage[noend]{algpseudocode}
\usepackage[linesnumbered,ruled,vlined]{algorithm2e}

\usepackage{colortbl}
\usepackage{bbm}
\usepackage{multirow}
\usepackage{threeparttable}
\usepackage{hhline}

\usepackage{graphics} % for EPS, load graphicx instead 
\usepackage{graphicx}
\usepackage{float}
\usepackage{caption} 
\usepackage{subcaption}
\usepackage{apxproof}  % proof sketch
\usepackage{xspace}

\newenvironment{myquote}
  {\begin{quote}
  \setlength{\parindent}{0pt}\setlength{\leftskip}{-1.5em}\setlength{\rightskip}{-2em}
  \small}
  {\end{quote}}

\newtheoremstyle{tight}
  {0.3em}   % ⬆️ 定理前的垂直间距
  {0.1em}   % ⬇️ 定理后的垂直间距
  {\itshape}  % 正文字体
  {10pt}      % 缩进
  {\scshape} % 定理标题（加粗 + 小型大写）
  {.}     % 定理标题后缀标点
  { }     % 标题与正文之间的空格
  {}      % 自定义标题格式（留空即可）
\theoremstyle{tight}

\newtheorem{theorem}{Theorem}[section]

\newtheorem*{prf}{Proof}

\newtheorem{assumption}[theorem]{Assumption}

\definecolor{codegreen}{rgb}{0,0.6,0}
\definecolor{codegray}{rgb}{0.5,0.5,0.5}
\definecolor{codepurple}{rgb}{0.58,0,0.82}
\definecolor{backcolour}{rgb}{0.95,0.95,0.95}
\definecolor{codeblue}{rgb}{0.0,0.5,0.69}
\definecolor{codered}{rgb}{0.8,0.0,0.0}
\definecolor{chocolate}{rgb}{0.82,0.41,0.12}
\definecolor{refbrown}{rgb}{0.65,0.16,0.16}

\definecolor{midnightblue}{rgb}{0.1, 0.1, 0.44}
\definecolor{teal}{rgb}{0.0, 0.5, 0.5}
\definecolor{lightgray}{rgb}{0.83, 0.83, 0.83}
\definecolor{maroon}{rgb}{0.5, 0.0, 0.0}

\newcommand{\tech}[1]{ #1 }

\newcommand{\smallcode}[1]{{\small\texttt{#1}}}

\newcommand{\hide}[1]{}

\lstdefinestyle{sqlstyle}{
  language = SQL, % SQL, TeX, Pseudocode
  morekeywords={UDF, OPERATE},
  keywordstyle=[2]{\color{codepurple}},
  morekeywords=[2]{},
  keywordstyle=[3]{\color{codeblue}},
  morekeywords=[3]{},
  commentstyle=\color{gray},
  keywordstyle=\color{codegreen}\bfseries\scshape,
  numberstyle=\tiny\color{codegray},
  stringstyle=\color{codered},
  basicstyle=\ttfamily\footnotesize,
  breakatwhitespace=false,         
  breaklines=true,                 
  captionpos=b,                    
  keepspaces=true,                 
  numbers=none,                    
  numbersep=5pt,                  
  showspaces=false,                
  showstringspaces=false,
  showtabs=false,                  
  tabsize=2,
}

\newcommand\footnoteref[1]{\protected@xdef\@thefnmark{\ref{#1}}\@footnotemark}

\def\app#1#2{%
  \mathrel{%
    \setbox0=\hbox{$#1\sim$}%
    \setbox2=\hbox{%
      \rlap{\hbox{$#1\propto$}}%
      \lower1.1\ht0\box0%
    }%
    \raise0.25\ht2\box2%
  }%
}

\AtBeginDocument{%
  \providecommand\BibTeX{{%
    \normalfont B\kern-0.5em{\scshape i\kern-0.25em b}\kern-0.8em\TeX}}}

\SetCommentSty{mycommfont}
\SetKwInput{KwInput}{Input} % Set the Input
\SetKwInput{KwOutput}{Output} % set the Output

\newcommand{\MAINNAME}{{\textsc{ReDD}}\xspace}

\newcommand{\MAJORITY}{{\textsc{MV}}\xspace}
\newcommand{\CONFLICT}{{\textsc{CF}}\xspace}
\newcommand{\CONFORMAL}{{\textsc{SCAPE}}\xspace}
\newcommand{\HYBRID}{{\textsc{SCAPE-Hyb}}\xspace}

\newcommand{\SPIDER}{{\textsc{Spider}}\xspace}
\newcommand{\BIRD}{{\textsc{Bird}}\xspace}
\newcommand{\Galois}{{\textsc{Galois}}\xspace}
\newcommand{\KAGGLE}{{\textsc{Fortune}}\xspace}
\newcommand{\BBC}{{\textsc{Premier}}\xspace}

\newcommand{\FDA}{{\textsc{FDA}}\xspace}
\newcommand{\CUAD}{{\textsc{CUAD}}\xspace}

\begin{document}

\title{Relational Deep Dive: Error-Aware Queries Over Unstructured Data}

\author{Daren Chao}
\affiliation{
  \institution{University of Toronto}
  \city{Toronto}
  \country{Canada}
}
\email{drchao@cs.toronto.edu}

\author{Kaiwen Chen}
\affiliation{
  \institution{University of Toronto}
  \city{Toronto}
  \country{Canada}
}
\email{kckevin.chen@mail.utoronto.ca}

\author{Naiqing Guan}
\affiliation{
  \institution{University of Toronto}
  \city{Toronto}
  \country{Canada}
}
\email{naiqing.guan@mail.utoronto.ca}

\author{Nick Koudas}
\affiliation{
  \institution{University of Toronto}
  \city{Toronto}
  \country{Canada}
}
\email{koudas@cs.toronto.edu}

\begin{abstract}
Unstructured data is pervasive, but analytical queries demand structured representations, creating a significant extraction challenge. Existing methods like RAG lack schema awareness and struggle with cross-document alignment, leading to high error rates. We propose ReDD (Relational Deep Dive), a framework that dynamically discovers query-specific schemas, populates relational tables, and ensures error-aware extraction with provable guarantees.
ReDD features a two-stage pipeline: (1) Iterative Schema Discovery (ISD) identifies minimal, joinable schemas tailored to each query, and (2) Tabular Data Population (TDP) extracts and corrects data using lightweight classifiers trained on LLM hidden states. A main contribution of ReDD is SCAPE, a statistically calibrated method for error detection with coverage guarantees, and SCAPE-HYB, a hybrid approach that optimizes the trade-off between accuracy and human correction costs.
Experiments across diverse datasets demonstrate ReDD's effectiveness, reducing data extraction errors from up to 30\% to below 1\% while maintaining high schema completeness (100\% recall) and precision. ReDD's modular design enables fine-grained control over accuracy-cost trade-offs, making it a robust solution for high-stakes analytical queries over unstructured corpora. 
\end{abstract}

\maketitle

\section{Introduction}

\begin{figure*}[t]
  \captionsetup{font=small, skip=4pt}
  \centering
  \captionsetup{skip=4pt}  % font=small, 
  \includegraphics[width=0.98\linewidth]{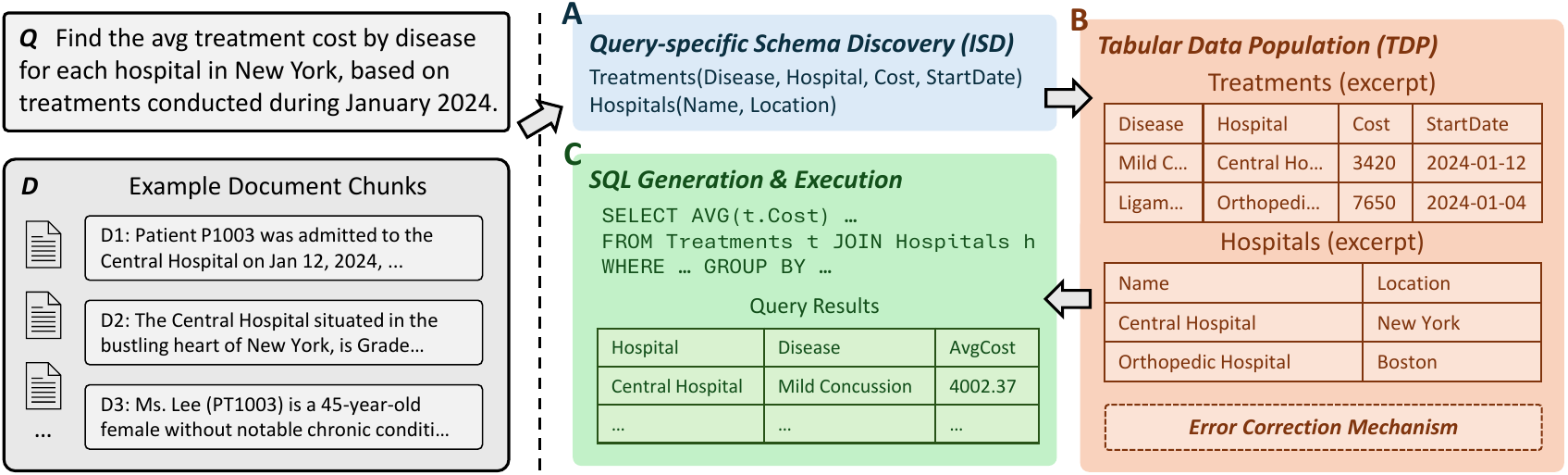}
  \caption{Overview of the query processing pipeline in \MAINNAME. 
    \textnormal{The left side of the dashed line shows the raw input, consisting of a natural language query and a collection of unstructured document chunks. The right side illustrates the core system workflow of \MAINNAME, comprising: (A) schema discovery; (B) data population; and (C) SQL query generation and execution (not the focus of this work). Within the data population component (B), an error correction mechanism is integrated to automatically detect and rectify low-confidence extractions, enabling controllable accuracy.}
  } 
  \label{fig:intro}
\end{figure*}

In many applications, including healthcare, finance, and engineering, the vast majority of the data produced is unstructured (in the form of reports, surveys, clinical trials, etc). Data analytics applications, however, in these domains require the data to be in a structured format. Consider, for example, analytical queries on the results of clinical trials for drug side effects or related queries in a financial domain to report on the top reasons identified for missed earnings of public companies in certain sectors. 
These queries may involve entity alignment, multi-hop reasoning, and statistical aggregation---tasks that are particularly difficult in the absence of structured representations. Structured data is also mandated by the strict accuracy requirements in these applications. As a result, unstructured data has to be processed to yield structured information for further downstream analytical processing. 

\noindent\textbf{Motivating Example.}
Consider the example in Figure~\ref{fig:intro}, which depicts a natural language query asking for the average treatment cost by disease for hospitals in New York in January 2024, alongside a collection of documents, depicted as document chunks for illustration purposes. These document chunks vary in focus: some describe treatment details (e.g., costs and diseases) like D1, others provide hospital metadata (e.g., location) like D2, and some contain patient profiles, irrelevant to the query, like D3. Answering this query requires aligning hospital names across document chunks and aggregating costs, a process complicated by the heterogeneous and fragmented nature of the data. Moreover, the absence of schemas or explicit semantic information (e.g., entity relationships) further hinders query answering. These challenges extend across domains-beyond medical reports to financial filings, legal documents, and more. Answering query $Q$ in Figure~\ref{fig:intro} directly from the data in the document collection is challenging.

Current methods, such as retrieval-augmented generation (RAG) \cite{lewis2021retrievalaugmentedgenerationknowledgeintensivenlp}, based on large language models (LLMs), attempt to answer queries over unstructured data by retrieving top-ranked documents based on query similarity and generating a response conditioned on a limited context. This design makes RAG optimized for precision, but offers limited control over recall---a property that is often more critical in database-style queries, such as those involving statistical aggregation and other analytical tasks, across documents as in Figure~\ref{fig:intro}. Moreover, RAG lacks schema awareness and struggles with cross-document entity alignment, such as linking hospital names across documents \cite{maekawa2025holistic, roberts2025needle}. 
State-of-the-art approaches for text-to-SQL \cite{10.1145/3709719} or traditional information extraction techniques \cite{manning08retrieval}, rely on predefined schemas or explicit semantic information, which is generally unavailable or undefined in unstructured corpora. 

In extracting value out of unstructured data, frameworks such as {\em DeepResearch} \cite{jones2025deepresearch,openai2025deepresearch,google2025deepresearch} and {\em DeepSearch} \cite{xai2024grok3} are gaining popularity. These are agentic frameworks that scan web pages (or documents) in a query-driven manner, producing detailed summaries and analysis in response to a {\em specific} user query. Motivated by such frameworks,  we seek to analyze semantic information from unstructured documents and extract {\em query-specific structured data}. Moreover, given that our focus is on running analytical queries on extracted structured data, we are interested in providing {\em error-aware query processing} with controllable accuracy---such as the ability to specify query-time error bounds or adjust accuracy-cost trade-offs. 

\vspace{0.3em}
\noindent\textbf{Relational Deep Dive (\MAINNAME) Overview.}
In this paper, we present \MAINNAME, (pronounced `ready') our proposed {\bf Re}lational {\bf D}eep {\bf D}ive framework. We aim for fine-grained accuracy control for the specified query, bridging the gap between raw documents and structured query execution. \MAINNAME integrates schema discovery (if applicable), structured data population, with error guarantees, and result synthesis in a cohesive pipeline {\em driven by the input query}. 

The pipeline begins with Iterative Schema Discovery (ISD), which processes the collection of documents at a suitable granularity (referred to as document chunks), iteratively refines a candidate set of tables and attributes, given a query $Q$, as more document chunks are processed. This procedure identifies the minimal schema required to answer $Q$, and uncovers latent semantic structures such as shared entities (join keys) across document chunks. For example, in Figure~\ref{fig:intro}(A), ISD discovers two tables---Treatments and Hospitals---along with attributes both directly relevant to the query (e.g., treatment cost and disease) and indirectly necessary for alignment (e.g., hospital name), even if not mentioned in the query itself. 
Once the schema is in place, \MAINNAME proceeds to Tabular Data Population (TDP). Each document chunk is parsed and converted into one or more rows across the discovered tables, depending on its content and relevance to the query $Q$. As exemplified in Figure~\ref{fig:intro}(B), chunk D1 populates the first row of the Treatments table, while D2 contributes a row to Hospitals. In contrast, D3 is irrelevant to the query and therefore does not populate any table. 
However, due to the ambiguity of natural language and the inherent variability of LLM outputs, extraction errors remain common---especially when attributes are implicit, phrased inconsistently, or missing altogether. To combat these challenges, we introduce a downstream correction module that proactively detects and corrects extraction errors, and supports human-in-the-loop intervention when required.

To enable this functionality, \MAINNAME supports controlled accuracy. Rather than treating the LLM as an infallible oracle, \MAINNAME quantifies extraction uncertainty and selectively corrects low-confidence outputs through lightweight classifiers trained on LLM hidden states and ensemble strategies. Specifically, \MAINNAME exploits the hidden representations (i.e., intermediate layer states) of LLMs, which encode rich contextual signals. These hidden representations are used as features for a suite of classifiers that predict extraction correctness---e.g., incorrect attribute values, wrongly assigned tables, or missing entries. 
These classifiers underpin a set of correction strategies, including ensemble agreement checks, conformal prediction, and fallback re-extraction. When classifier disagreement or uncertainty is high, we either reprocess the document chunk or flag it for human-in-the-loop review. This modular design enables fine-grained control over the accuracy-efficiency trade-off at query time as we detail in \S\ref{sec:hybrid}. 

In our experiments across multiple domains, we observe that direct data extraction by LLMs can yield error rates (measured as the proportion of incorrectly populated rows) of up to 30\% on certain challenging datasets. With \MAINNAME's correction techniques, these error rates consistently drop below 1\%, without requiring schema supervision or domain-specific rule engineering. These gains underscore the effectiveness of our end-to-end framework in achieving not only high accuracy but also controllable extraction quality, even over large-scale unstructured document collections. 
This paper makes the following contributions:

\begin{itemize}[itemsep=0pt, topsep=0pt, leftmargin=10pt]
\item We present \MAINNAME, a query-specific framework for structured query execution over unstructured text. The framework bridges the gap between raw documents and structured query processing by dynamically discovering schemas, populating tables, and correcting extraction errors. \MAINNAME achieves this through a cohesive pipeline comprising Iterative Schema Discovery (ISD), Tabular Data Population (TDP), and error detection and control, enabling accurate query answering with structured representations even in the absence of predefined schemas or annotations.

\item We develop a two-stage schema discovery pipeline that constructs minimal, joinable tables tailored to each query.
Empirical results demonstrate that this two-phase approach yields more accurate and query-complete schemas compared to single-phase alternatives.

\item We introduce \CONFORMAL (Spatial Conformal Activation Partitioning for Errors) a statistically calibrated method that guarantees error coverage (Theorem \ref{thm:scape_coverage}) while asymptotically being optimal in minimizing human correction costs (Theorem \ref{thm:optimality}) by partitioning high-dimensional non-conformity scores (Algorithm 1).
Moreover, we introduce \HYBRID, a hybrid approach that integrates conflict-aware signals with \CONFORMAL, maintaining its well calibrated properties (Theorem \ref{thm:scape_coverage}) while enabling flexible trade-offs between error detection recall and correction costs with provable guarantees (Theorem \ref{thm:scape_hyb_advantage}).

\item We validate \MAINNAME on several real-world datasets, showing that it scales to large unstructured document collections and reduces query error rates from up to 30\% to below 1\%.
The results underscore \MAINNAME's effectiveness in transforming unstructured text into query-ready structured data with provable guarantees.
\end{itemize}

\noindent
In \S\ref{sec:framework} we introduce the core components of the \MAINNAME framework. 
\S\ref{sec:datapop}-\ref{sec:hybrid} present our error management module for data extraction, followed by \S\ref{sec:schema} presenting our schema discovery methodology. \S\ref{sec:exp} reports experimental results across multiple datasets\footnote{Our code and datasets are available at: \url{https://github.com/daren996/ReDD}.}. 
\S\ref{sec:related} reviews related work and we present our closing
remarks in \S\ref{sec:con}.
%We discuss practical considerations and broader implications in \S\ref{sec:discuss}, and conclude in \S\ref{sec:con}.

\section{The \MAINNAME Framework}
\label{sec:framework}

\MAINNAME proceeds in two stages and transforms unstructured textual data into query-ready structured tables. The system accepts as input a natural language query $q$ and a collection of unstructured documents. These documents are segmented into semantically coherent \textit{chunks}, denoted as $D = \{d_1, d_2, \ldots, d_n\}$, where each chunk $d_i$ is a contiguous span of text bounded by semantic discontinuities (e.g., paragraph breaks). 
We treat each chunk as the minimal semantic unit of processing. Depending on its content, a chunk may yield one or multiple rows\footnote{
  For brevity and to ease presentation, in the remainder of this section we assume that each chunk yields one row for illustration; extending to other scenarios such as one-to-many and many-to-one mappings is straightforward (\S\ref{sec:datapop}) and imposes no changes to the downstream modules. We empirically validate these along with their trade-offs, in \S\ref{sec:exp_population_onetomany} (one-to-many) and \S\ref{sec:exp_population_acc} (many-to-one).
} distributed across one or more (initially unknown) tables. The goal of \MAINNAME is to (i) discover the latent schema $S_D^q$ required to answer $q$, and (ii) populate that schema with tuples extracted from $D$, supporting accurate and error-aware query execution. 

To answer complex analytical queries over unstructured text, it is often necessary to recover latent structured representations---namely, relational tables and their schemas---that are not explicitly present in the input. In some cases (as exemplified by systems such as Galois \cite{satriani2025logical} and Palimpzest \cite{liu2025palimpzest}), the schema
may be known or provided (see \S\ref{sec:related} for details). Although \MAINNAME handles this case naturally, we take a more general and principled approach by enabling automated schema discovery. 
This process is complicated by several factors:
\begin{itemize}[itemsep=0pt, topsep=0pt, leftmargin=10pt]
  \item The number of tables, their schemas, and the mapping from document chunks to tables are unknown a priori. 
  \item The query $q$ may require aggregating information distributed across multiple latent tables to produce a complete answer.
\end{itemize}
This setting reflects real-world scenarios where no external schema, entity linking, or domain-specific annotations are available, and the query-specific structured tables must be discovered dynamically from text.

\MAINNAME addresses these challenges through its two-stage pipeline: Iterative Schema Discovery (ISD) and Tabular Data Population (TDP), which includes an error correction module (see \S\ref{sec:datapop}-\ref{sec:hybrid}). The entire pipeline is driven by the input query $q$, while also mining latent semantics across chunks to construct accurate, structured outputs. Given a query specific schema $S_D^q$ at hand (a process described in \S\ref{sec:schema}) we next detail the TDP phase (\S\ref{sec:datapop}-\ref{sec:hybrid}) followed by ISD.

\section{Tabular Data Population}
\label{sec:datapop}

Given a query-specific schema $S_D^q$, the Tabular Data Population (TDP) stage inserts tuples into the schema by extracting structured data from document chunks. 
Each chunk $d_k$ is independently processed to yield structured rows aligned with the schema semantics. The TDP stage follows a fixed, two-step pipeline, in which lightweight, LLM-guided prompt functions extract and format relevant data\footnote{\tech{Concrete prompt templates and implementation details are provided in Appendix~\ref{apx:prompt_datapop}.}}. 
\begin{itemize}[itemsep=0pt, topsep=0pt, leftmargin=10pt]
  \item 
  Table Resolver $\mathcal{X}_T$.
  For every chunk $d_k$, the resolver selects the most semantically compatible table $T_k \in S_D^q$: 
  \begin{equation}
    T_k = \mathcal{X}_T\left( d_k, S_D^q \right), \quad \text{for } k = 1..n 
    \label{eq:table_rsl}
  \end{equation}
  where $T_k$ denotes the identifier of the selected target table.
  \item 
  Attribute Extractor $\mathcal{X}_A$.
  Given the pair $(d_k, T_k)$, the extractor iteratively fills each attribute $a_i \in T_k$:
  \begin{equation}
    v_{k,i} = \mathcal{X}_A\left( d_k, a_i, T_k \right), \quad \text{for } k = 1..n,\ a_i \in T_k
    \label{eq:attr_ext}
  \end{equation}
  where $v_{k,i}$ denotes the value of attribute $a_i$ extracted from chunk $d_k$. The resulting tuple $(v_{k,1}, \ldots, v_{k,m})$ forms a single row to be inserted into table $T_k$.
\end{itemize}
This procedure extracts, for each chunk, a semantically aligned table and a complete attribute-value tuple, yielding one structured row per chunk as the final output of data population. It also generalizes to one-to-many settings: the table resolver can return multiple candidate tables per chunk, and the attribute extractor can extract multiple tuples per table accordingly.

Due to the inherent ambiguity of natural language and the stochastic behavior of LLM outputs, the chunk-wise extraction of attribute values can lead to extraction inconsistencies and errors, such as incorrect or incomplete attribute values, or even mis-assigned rows. 
Relying solely on an LLM in TDP during table population yields error rates (measured as the proportion of incorrectly populated rows) of up to 30\% on challenging datasets (see \S\ref{sec:exp_population}). 
%The ISD stage, by contrast, depends on the entire document collection and incrementally maintains a global schema state, refining it through iterative integration of information across chunks. 
The TDP stage, makes local decisions: it processes each document chunk during data population, one at a time, mapping it to one or more tables and extracting tuple(s). Lacking a holistic view, TDP cannot revise earlier decisions based on broader context, making tuple extraction inherently error-prone. This limitation instigates the introduction of additional mechanisms to mitigate such errors.

\subsection{Error Detection Approaches}
\label{sec:baselines}

While Tabular Data Population (TDP) extracts tuples and populates query table(s) whose schema is discovered for a query $q$ during ISD, it may introduce extraction errors due to the inherent uncertainty of LLM outputs. 
In high-stakes settings, even a small number of erroneous entries may lead to incorrect conclusions in downstream analyses. 
\MAINNAME operation is geared towards a specific query $q$ and it includes mechanisms to identify and correct data extraction errors, ensuring that each relational table entry is grounded in the source document. 
Central to our approach towards error-free data extraction is assessing the trustworthiness of each extracted value, utilizing small, open-weight LLMs. Their compact size allows for local deployment, while being open-weight enables the development of specialized, tunable algorithms around them. Based on this assessment, we decide when to trigger corrective actions and/or abstain from extracting the value and trigger a human review to assist our algorithms with extraction. Thus, \MAINNAME incorporates a human-in-the-loop as a first-class primitive. 

Before TDP commences, we construct a small labeled dataset, denoted as $\mathcal{D}_{\text{cls}}$, which serves as training data for lightweight classifiers that predict the correctness of extracted values during TDP. 
Each entry in $\mathcal{D}_{\text{cls}}$ is generated by applying the LLM prompts and using the same LLM model (denoted as $M^{\text{TDP}}$) as in the TDP stage (as defined in Equations~(\ref{eq:table_rsl})-(\ref{eq:attr_ext})) to a limited set of document chunks, producing candidate table rows. 
The ground-truth labels for these entries can be obtained through one of the following two approaches: 
\begin{enumerate}[itemsep=0pt, topsep=0pt, leftmargin=15pt]
    \item \textbf{Human-Verified Labeling}: A user manually verifies each extracted value against the source text, annotating whether it is correct or incorrect.
    \item \textbf{LLM-Committee-Based Labeling}: A committee of powerful LLMs (e.g., OpenAI-o3 \cite{openai_o3_and_o4mini} and Claude-4 \cite{anthropic_claude_4}) independently extracts and evaluates the same tuples. 
    % The correctness of a candidate extraction is determined by consensus: if the initial LLM's output disagrees with the committee's assessment, the committee's label is taken as ground truth. 
    The correctness of a candidate extraction is determined by the consensus of all committee members: if the initial LLM's output disagrees with the committee's assessment, the committee's label is taken as the ground truth, and the extraction is marked as incorrect. 
\end{enumerate}
\noindent 
Formally, let $\mathcal{M} {=} \{M_1, \dots, M_K\}$ denote the committee of $K$ powerful LLMs. For a candidate tuple $t$ extracted by the initial LLM $M^{\text{TDP}}$, the label $y_t$ is assigned as: 
\begin{equation}
\small
  y_t = 
  \begin{cases} 
    1 & \text{if } t \text{ matches majority of } \mathcal{M} \text{ or human confirms,} \\
    0 & \text{otherwise.} 
  \end{cases}
  \label{eq:label_assignment}
\end{equation}
This hybrid approach ensures high-quality labeled data while minimizing reliance on manual annotation. The required size of $\mathcal{D}_{\text{cls}}$ is small, as we demonstrate in \S\ref{sec:exp_population} by varying its size and measuring its impact on extraction accuracy.
These classifiers (\S\ref{sec:latent}) enable error detection during the extraction process.

Following error detection, \MAINNAME applies error resolution measures, such as committee-based inference using more powerful LLMs, or escalation to human verification for low-confidence cases. 
Such error resolution methods incur additional cost (e.g., monetary, time burden, etc), and need to be minimized. We refer to this overhead as {\em human correction cost} or {\em cost} interchangeably. This cost is proportional to the number of entries flagged as erroneous, i.e., those with predicted error label $\hat{y} {=} 1$. 
Notice that in case error detection instigates {\em false positives} ($\hat{y} {=} 1$ but the true label $y {=} 0$), this adds extra human correction cost (such as human validation of already correct extractions). Thus, although correct error detection is important (to minimize false negatives), reducing false positives is a major design requirement as well.

Our approach deploys a collection of lightweight classifiers utilizing the hidden states of the LLM ($M^{\text{TDP}}$) to quantify the correctness of each token output. We will start by introducing notation and two baseline methods. The first, \MAJORITY, utilizes majority voting over the binary outputs of individual classifiers to quantify token-level correctness. The second, \CONFLICT, is a conservative refinement of \MAJORITY designed to identify strong disagreement among classifier predictions, thereby reducing false positives. We then introduce our main proposals \CONFORMAL and \HYBRID in \S\ref{sec:hybrid}. 

\subsection{Error Detection via Latent Representations}
\label{sec:latent}

A cornerstone of our approach is a lightweight binary classifier that determines whether each extracted attribute value is correct, using the LLM's own hidden representations. Recent studies have demonstrated that LLMs' internal states encode rich information about the truthfulness of their outputs \cite{orgad2024llms}, which we exploit to identify potential extraction errors in \MAINNAME.

\vspace{0.3em}
\noindent\textbf{Leveraging LLM Hidden States.}
As described in \S\ref{sec:datapop}, for each document chunk $d_k$, the TDP stage first assigns a target table $T_k$ via the table resolver $\mathcal{X}_T$ (Equation~(\ref{eq:table_rsl})), and then extracts attribute values $v_{k,i}$ using the attribute extractor $\mathcal{X}_A$ (Equation~(\ref{eq:attr_ext})). 
During these steps, we have full access to the hidden states of the LLM $M^{\text{TDP}}$. All LLM outputs (including the assigned table name $T_k$ and each extracted attribute value $v_{k,i}$) are generated as sequences of output tokens. 
Let $\mathbf{w}_{k,\text{table}} = (w_{k,\text{table}}^{1}, \ldots, w_{k,\text{table}}^{m})$ be the token sequence corresponding to the assigned table name $T_k$, and let $\mathbf{w}_{k,\text{attr-}i} = (w_{k,\text{attr-}i}^{1}, \ldots, w_{k,\text{attr-}i}^{m_i})$ be the token sequence corresponding to the extracted value $v_{k,i}$ for attribute $a_i$ (the $i$-th attribute) in $T_k$. 
Let $h^{(l)}(w)$ denote the hidden state at layer $l$ corresponding to a token $w$. 
To obtain a compact representation of the LLM extraction outputs (for both table names and attribute values), we apply mean-max pooling across the token-level hidden states at each layer, then concatenate the pooled vectors: 
\begin{equation} 
\small
\begin{aligned}
  &h_{k,\text{table}}^{(l)} = \texttt{concat}\left( \underset{j}{\text{mean}}\ h^{(l)}(w_{k,\text{table}}^{j}), \max_j h^{(l)}(w_{k,\text{table}}^{j}) \right), \\
  &h_{k,\text{attr-}i}^{(l)} = \texttt{concat}\left( \underset{j}{\text{mean}}\ h^{(l)}(w_{k,\text{attr-}i}^{j}), \max_j h^{(l)}(w_{k,\text{attr-}i}^{j}) \right),
\end{aligned} \label{eq:min_max} \end{equation}
where $h_{k,\text{table}}^{(l)} {\in} \mathbb{R}^{2d}$ and $h_{k,\text{attr-}i}^{(l)} {\in} \mathbb{R}^{2d}$ denote the layer-$l$ hidden representations for table name $T_k$ and attribute value $v_{k,i}$, respectively. 
Here, $d$ is the size of the hidden state of the underlying LLM $M^{\text{TDP}}$. 
Since both vectors are computed through identical mean-max pooling operations, share the same dimensionality, and follow the same procedure to train the per-layer classifiers (introduced in \S\ref{sec:mv_cf}), we adopt a unified notation for simplicity. 
Specifically, we denote each concatenated hidden representation as $h_{k,\circ}^{(l)}$, where $\circ$ stands for either extracted table names or attribute values. This simplification relaxes notation without loss of generality. 
In addition, we use $y_{k,\circ}$ to refer to the ground truth label of each extraction, indicating whether the extracted item (table or attribute value) is erroneous or not. A value of $y_{k,\circ} {=} 1$ denotes an error, while $y_{k,\circ} {=} 0$ indicates correctness.

Let $\mathcal{L}$ denote the set of LLM layers from which hidden states are extracted. This set may include all layers or a selected subset\footnote{When a subset of layers is used to train classifiers for error prediction, the index $l$ refers to the $l$-th element in $\mathcal{L}$, not the original layer number in the LLM.}, as recent studies have shown that certain layers encode richer and more informative signals than others \cite{jawahar2019does,meng2022locating}. 
The aggregated hidden representations obtained using Equation~(\ref{eq:min_max}) are subsequently used as input features for binary classifiers that predict whether the corresponding table assignment or extracted attribute value is likely to be incorrect.
These hidden states capture rich contextual and semantic cues from both the document and the query, making them an informative signal for spotting inconsistencies or mistakes in the LLM's own outputs. The classifiers are trained using $\mathcal{D}_{\text{cls}}$.

\vspace{0.3em}
\noindent\textbf{Example.} 
Consider the value ``Central Hospital'' extracted for the attribute \textit{Name} in the table \textit{Hospitals}. The LLM may tokenize this value into multiple subword tokens as the output, such as ``Central'' $w_{k,\text{attr-}i}^1$ and ``Hospital'' $w_{k,\text{attr-}i}^2$. For each LLM layer $l$, we obtain the corresponding hidden states $h^{(l)}(w_{k,\text{attr-}i}^1)$ and $h^{(l)}(w_{k,\text{attr-}i}^2)$ for all tokens in the value. We then compute both the mean and max over these token representations and concatenate the results to obtain a fixed-length vector $h_{k,\text{attr-}i}^{(l)}$ following Equation~(\ref{eq:min_max}). The resulting vectors ${h_{k,\text{attr-}i}^{(l)}} \mid l {\in} \mathcal{L}$ are used as input to the per-layer binary classifiers that estimate whether the extracted value is erroneous. 

\subsection{Voting Based Methods: \MAJORITY and \CONFLICT}
\label{sec:mv_cf}

To detect extraction errors from TDP, we begin with a straightforward but effective baseline: performing majority voting on predictions from classifiers trained on hidden states from each layer. 

\vspace{0.3em}
\noindent\textbf{Layer-wise Error Classifiers.}
For each LLM layer $l {\in} \mathcal{L}$, we train a lightweight binary classifier\footnote{Each classifier is a multilayer perceptron (MLP) with a sigmoid output.} $f^{(l)}{:} \mathbb{R}^{2d} {\rightarrow} [0,1]$ to predict if the extraction associated with hidden state $h_{k,\circ}^{(l)}$ (as in \S\ref{sec:latent}) is erroneous: 
\begin{equation} 
\begin{aligned}
  \pi_{k,\circ}^{(l)} = f^{(l)} \left(h_{k,\circ}^{(l)}\right),
\end{aligned} \label{eq:pi} \end{equation}
where $\pi_{k,\circ}^{(l)}$ is the predicted probability of an error. The classifier per layer is trained independently using binary cross-entropy loss.
To convert probabilities into binary decisions, a fixed threshold $\theta$ is applied:
\begin{equation}
\hat{y}_{k,\circ}^{(l)} = \mathbf{1} \left[ \pi_{k,\circ}^{(l)} > \theta \right]. 
\label{eq:haty} \end{equation}
Here, $\theta$ determines the decision boundary of each classifier. 
Its choice is typically set arbitrarily (e.g., 0.5) and ignores classifier-specific mis-calibrations, as well as classifier dependencies. 

\vspace{0.3em}
\noindent\textbf{Majority Voting (\MAJORITY).}
\MAJORITY aggregates the binary predictions $\hat{y}_{k,\circ}^{(l)}$ ($l\in\mathcal{L}$) via majority voting. The idea is that if most classifiers agree an extraction is wrong, it's likely to be so:
\begin{equation} 
\hat{y}_{k,\circ}^{\MAJORITY} = \mathbf{1}\left[\sum_{l\in\mathcal{L}} \hat{y}_{k,\circ}^{(l)} > \frac{|\mathcal{L}|}{2} \right].
\label{eq:error_mv} \end{equation}
This simple rule effectively denoises isolated errors from individual classifiers by requiring consensus across layers. However, majority voting offers no explicit control over the false positive and false negative rates, making it hard to balance detection accuracy and correction cost as application demands vary.

\vspace{0.3em}
\noindent\textbf{Conflict Filtering (\CONFLICT).}
While \MAJORITY relies on agreement, classifier disagreement can also be informative. \CONFLICT builds on this idea by measuring how much conflict exists among the layer-wise predictions. Intuitively, if different layers disagree about whether an extraction is erroneous, it likely reflects ambiguity or model hesitation.
We define the conflict score $\kappa$ as the number of classifiers that disagree with the majority vote:
\begin{equation} 
  \kappa = \sum_{l\in\mathcal{L}} \mathbf{1}\left\{\hat{y}_{k,\circ}^{(l)} \ne \hat{y}_{k,\circ}^{\MAJORITY}\right\},
\label{eq:error_cf_kappa} 
\end{equation}
An extraction is flagged as potentially erroneous if the conflict score exceeds a tunable threshold $\tau^{\CONFLICT}$: 
\begin{equation*} 
\begin{aligned}
\hat{y}_{k,\circ}^{\CONFLICT} = 
  \begin{cases}
    \hat{y}_{k,\circ}^{\MAJORITY}, & \text{if } \kappa < \tau^{\CONFLICT} \\
    1, & \text{if } \kappa \geq \tau^{\CONFLICT} \\
  \end{cases}
\end{aligned} \end{equation*}
Empirically, a higher $\kappa$ value signifies a high-conflict case and often corresponds to true errors, making \CONFLICT an effective strategy for reducing false negatives. In contrast, a small $\kappa$ value signifies low disagreement and points to more confident decisions. 
\hide{
Uniformly confident but incorrect predictions (where all layers agree yet the result is wrong) are relatively rare (see \S\ref{sec:}). {\color{red} do we comment on it anywhere?}
}
The threshold $\tau^{\CONFLICT}$ controls the sensitivity of the \CONFLICT strategy: a smaller value flags potential errors even in low-conflict cases, thereby improving recall by capturing more true errors than \MAJORITY, but at the cost of increased false positives and correction overhead. 
% Empirically, when $\tau^{\CONFLICT}$ is large, a higher $\kappa$ value signifies a high-conflict case often corresponding to true errors, making \CONFLICT an effective strategy for reducing false negatives. In contrast, a small $\kappa$ value signifies low disagreement and points to more confident decisions. Uniformly confident but incorrect predictions (where all layers agree yet are wrong) are relatively rare (see \S\ref{sec:}). Compared to \MAJORITY, \CONFLICT improves recall---capturing more true errors---by leveraging inter-layer disagreement, but at the cost of increased false positives and correction overhead. 

\vspace{0.3em}
\noindent\textbf{Limitations.}
Both methods lack a principled way to control the trade-off between detection accuracy and correction cost. While \CONFLICT introduces a tunable conflict threshold $\tau^{\CONFLICT}$, its impact on accuracy and cost is heuristic, with no calibrated semantics or statistical guarantees. This motivates the development of a more controlled algorithm with formal error-rate guarantees to be introduced next.

\section{Enabling Error Detection Tradeoffs}
\label{sec:hybrid}

To address the limitations of the previous methods and enhance control over the accuracy of error detection and associated error correction costs, we present two proposals, \CONFORMAL and a hybrid approach \HYBRID. Instead of relying on
binary classifier predictions and aggregations thereof to quantify the accuracy of a prediction, our proposals leverage the continuous activation probabilities of classifiers jointly to quantify uncertainty more precisely. 
\begin{itemize}[itemsep=0pt, topsep=0pt, leftmargin=10pt]
\item \CONFORMAL (\S\ref{sec:scape}): 
  is a statistically calibrated method that quantifies prediction uncertainty. Its aim is to reduce false negatives (and thus undetected errors) but may increase false positives in the process and thus increase cost (i.e., imposing extra human labour to check the result of the extraction). It allows adjustment of correction aggressiveness via a coverage parameter $\alpha$.
\item \HYBRID (\S\ref{sec:hyb}): 
  enhances \CONFORMAL with \CONFLICT, allowing a more flexible balance between accuracy and human correction cost. 
\end{itemize}
In essence, these algorithms enable a tradeoff between prediction accuracy for erroneous data extractions by the LLM and (human) correction cost, by introducing two key parameters: 
\begin{itemize}[itemsep=0pt, topsep=0pt, leftmargin=10pt]
\item Coverage Threshold $\alpha$: 
  Higher values reduce false negatives (undetected errors) but may increase false positives, leading to higher correction costs.
\item Conflict Weight $\lambda$: 
  Controls how strongly the conflict-aware algorithm \CONFLICT influences correction decisions. A higher $\lambda$ gives greater weight to \CONFLICT, encourages more conservative decisions, increasing the chance of flagging potential errors (thus reducing false negatives), but may also lead to more cautious behavior and a rise in false positives.
\end{itemize}
These techniques enable precise and cost-aware control during data extraction. This is essential in applications where undetected errors are unacceptable. 

\subsection{\CONFORMAL: Spatial Conformal Activation Partitioning for Errors}
\label{sec:scape}

The \CONFORMAL framework introduces a novel approach to uncertainty quantification by leveraging a high-dimensional non-conformity score space \cite{vovk2005algorithmic, barber2023conformal}, contrasting with the previously introduced methods that rely on independent thresholds for binary classifiers \cite{shafer2008tutorial, barber2023conformalpredictionexchangeability, angelopoulos2022gentleintroductionconformalprediction}. 
Instead of treating each classifier's decision boundary in isolation, this technique constructs a multi-dimensional (spatial) score by combining outputs from classifiers trained on different layers. 
The key innovation lies in partitioning this high-dimensional space into adaptive cells centered around calibration data, which are then ranked by the empirical ratio of correct to incorrect labels observed among the calibration samples contained in each cell. 

By selecting regions with low concentrations of incorrect labels (where the classifiers historically perform well), the method generates more efficient and precise prediction sets while maintaining guaranteed coverage.
This approach avoids rigid thresholding or weighted aggregation, instead exploiting the richer geometric structure of the multi-dimensional space to better separate true from false predictions, yielding smaller and more informative uncertainty sets. 
The framework's flexibility allows it to outperform traditional conformal prediction, particularly in scenarios where classifiers provide complementary information across input or output regions.
\CONFORMAL enables coverage (or recall) control of error detection under mild distributional assumptions, ensuring a specified proportion of true errors is identified with high probability.

\vspace{0.3em}
\noindent\textbf{Non-Conformity Vectors.}
For each extracted item (either a table name or an attribute value) with hidden representations $\{h_{k,\circ}^{(l)}\}_{l\in\mathcal{L}}$ (as per \S\ref{sec:latent}), we first obtain the sigmoid outputs $\pi_{k,\circ}^{(l)} = f^{(l)}(h_{k,\circ}^{(l)})\in[0,1]$ from each layer-wise classifier $f^{(l)}$ at layer $l{\in}\mathcal{L}$ (as detailed in Equation~(\ref{eq:pi})). 
We then define a \emph{multi-dimensional non-conformity vector} $\mathbf{s}(c) {\in} \mathbb{R}^{|\mathcal{L}|}$ for each candidate label $c{\in} \{0,1\}$ (where $c{=}1$ denotesw erroneous extraction and $c{=}0$ denotes correct extraction) as: 
\begin{equation} 
\begin{aligned}
  \mathbf{s}(c) = \Bigl[ s_l(c) \Bigr]_{l=1}^{\mathcal{L}}, \quad
  \scalebox{0.8}{$\displaystyle
    \text{where } s_l(c) =
    \begin{cases}
    1 - \pi_{k,\circ}^{(l)} & \text{if } c = 1, \\
    \pi_{k,\circ}^{(l)}     & \text{if } c = 0.
    \end{cases} 
  $}
\end{aligned} \label{eq:nc_score} \end{equation}
which reflects how atypical the outputs of the layer-specific error classifiers are, under the assumption that the extraction is either erroneous or correct. 

\vspace{0.3em}
\noindent\textbf{Cell Construction and Selection.}
We leverage a labeled dataset $\mathcal{D}_{\text{cal-base}} = \{(x_i, y_i)\}_{i=1}^{N_\text{cal-base}}$, constructed using the same procedure as in \S\ref{sec:baselines}, where each $x_i$ denotes an extracted item and $y_i {\in} \{0,1\}$ indicates whether the extraction is erroneous (1) or correct (0).\footnote{$\mathcal{D}_{\text{cal-base}}$ is a small, user-curated labeled dataset, generated by applying the LLM prompts used in TDP stage (as Equations (\ref{eq:table_rsl})-(\ref{eq:attr_ext})) to a small number of document chunks. The dataset size is varied in the experiments to study its impact on overall accuracy in \S\ref{sec:exp_population}.} We randomly split this dataset into two disjoint subsets, $\mathcal{D}_{\text{cell}}$ and $\mathcal{D}_{\text{re-cal}}$, which serve distinct purposes in the calibration procedure. 
\begin{itemize}[itemsep=0pt, topsep=0pt, leftmargin=10pt]
  \item $\mathcal{D}_{\text{cell}} = \{(x_i,y_i)\}_{i=1}^{N_\text{cell}}$ used to construct cells in the non-conformity score space by applying $k$-means clustering to the score vectors $\mathbf{s}(\hat{y}_i)_{i=1}^{N_\text{cell}}$, producing $K$ clusters. Each cluster defines a cell, yielding a partition of the score space into non-overlapping regions $C_1, \ldots, C_K \subset \mathbb{R}^{|\mathcal{L}|}$, where any new score vector can be assigned to one of the cells by finding its nearest cluster centroid in Euclidean space. 
  \item $\mathcal{D}_{\text{re-cal}} = \{(x_j,y_j)\}_{j=1}^{N_\text{re-cal}}$ used to select cells for coverage. 
\end{itemize}
Each cell $C_m$ groups similar non-conformity patterns. 
To identify the most reliable regions of the score space for detecting true extraction errors, we rank cells based on their \emph{false-to-true ratio} on $\mathcal{D}_{\text{re-cal}}$, that is, for each cell $C_m$, and for $c {\in} \{0,1\}$ the number of examples with $y_j{=}c$ whose score vectors $\mathbf{s}(1{-}c)$ fall into the cell $C_m$ (i.e., false examples), divided by the number of examples $y_j{=}c$ whose $\mathbf{s}(c)$ also fall into the cell (i.e., true examples): 
\begin{equation}
  \rho_m = \frac{\sum_{c\in\{0, 1\}} |{(x_j,c) \in \mathcal{D}_{\text{re-cal}} : \mathbf{s}(1-c) \in C_m}|}{\sum_{c\in\{0, 1\}} |{(x_j,c) \in \mathcal{D}_{\text{re-cal}} : \mathbf{s}(c) \in C_m}|}. 
\label{eq:rho} \end{equation}
We rank the cells in ascending order of $\rho_m$, prioritizing those where non-error examples are least likely to be mistaken as errors.

To guarantee the desired coverage level, we select the smallest set of top-ranked cells such that the score vectors of the true labels for at least $\lceil (1 - \alpha)(N_{\text{re-cal}} + 1) \rceil$ examples in $\mathcal{D}_{\text{re-cal}}$ fall within the selected cells. 
Formally, let all cells be \emph{re-indexed} as $C{(1)}, C_{(2)}, \ldots, C_{(K)}$, sorted in order of increasing false-to-true ratio. 
We define the selected region as $\mathcal{C}_\alpha \subset \mathbb{R}^{|\mathcal{L}|}$: 
\begin{equation} 
\begin{aligned}
  \mathcal{C}_\alpha = \bigcup_{j=1}^{\eta^*} C_{(j)}, \quad
  \begin{array}{l}
    \scalebox{0.8}{$\displaystyle
      \text{where } \eta^* {=} \min \Bigl\{ \eta {\in} 1..K \mid \bigl| \bigl\{ (x_j, y_j) \in \mathcal{D}_{\text{re-cal}} : \bigr. \Bigr.
    $} \\
    \scalebox{0.8}{$\displaystyle
      \quad\quad\quad \Bigl. \bigr. \mathbf{s}(y_j) \in \mathcal{C}_\alpha \bigr\} \bigr| \geq \bigl\lceil (1 - \alpha)(N_{\text{re-cal}} + 1) \bigr\rceil \Bigr\}
    $}
  \end{array} 
  % \scalebox{0.8}{$\displaystyle
  %   \text{where } \eta^* {=} \min \Bigl\{ \eta {\in} \{1, \ldots, K\} \mid \Bigr.
  % $} \\
  % \scalebox{0.8}{$\displaystyle
  %   \Bigl. \left| \left\{ (x_j, y_j) \in \mathcal{D}_{\text{re-cal}} : \mathbf{s}(y_j) \in \mathcal{C}_\alpha \right\} \right| \geq \left\lceil (1 - \alpha)(N_{\text{re-cal}} + 1) \right\rceil \Bigr\}
  % $}
\end{aligned} \label{eq:c_alpha} \end{equation}
where $\alpha {\in} (0, 1)$ is the user-specified miscoverage tolerance.

\vspace{0.3em}
\noindent\textbf{Test-Time Inference.}
At test time, for each new extraction with hidden representations $\{h_{k,\circ}^{(l)}\}_{l\in\mathcal{L}}$ (as per \S\ref{sec:latent}), we compute the non-conformity vectors $\mathbf{s}(y)$ for both possible labels $y {\in} \{0,1\}$ (via Equation~(\ref{eq:nc_score})), and define the conformal prediction set as: 
\begin{equation}
  \hat{y}^{\CONFORMAL}_{k,\circ} = \left\{ y \in \{0,1\} \mid \mathbf{s}(y) \in \mathcal{C}_\alpha \right\}.
\label{eq:haty_scape} \end{equation}
By evaluating both candidate labels, we construct a prediction set that includes multiple labels only when necessary to satisfy the desired $\alpha$-coverage for error detection, while keeping the set as small as possible to reduce correction cost. 
If $\hat{y}^{\CONFORMAL}_{k,\circ}\!=\!\{0\}$, we accept the extraction as correct. Otherwise, if the prediction set contains 1 (i.e., $\{1\}$ or $\{0,1\}$), the extraction is flagged for potential error and triggers a correction step, typically by routing the value for human verification and correction. 

\CONFORMAL is presented as Algorithm~\ref{alg:scape}, where line~\ref{line:nc_score} corresponds to the computation of non-conformity vectors as defined in Equation~(\ref{eq:nc_score}), lines~\ref{line:cell}-\ref{line:rank} implement the clustering and ranking procedure described in Equation~(\ref{eq:rho}), line~\ref{line:top_cell} selects cells according to the coverage constraint in Equation~(\ref{eq:c_alpha}), and line~\ref{line:inference} defines the prediction set $\hat{y}^{\CONFORMAL}_{k,\circ}$ for each new extraction as Equation~(\ref{eq:haty_scape}).

{\small
\begin{algorithm}[h]
\captionsetup{font=small}
\caption{Spatial Conformal Activation Partitioning for Errors (\CONFORMAL)}
\label{alg:scape}
\begin{algorithmic}[1]
  \Require Calibration dataset $\mathcal{D}_{\text{cal-base}} = \{(x_i,y_i)\}_{i=1}^{N_\text{cal-base}}$, split into~$\mathcal{D}_{\text{cells}}$ and $\mathcal{D}_{\text{re-cal}}$; Coverage level $\alpha \in (0,1)$; 
  % \Statex Layer classifiers $\{f^{(l)}\}_{l \in \mathcal{L}}$ with sigmoid outputs $\pi^{(l)}(\cdot)$; 
  \Ensure Error prediction set $\hat{y}^{\CONFORMAL}_{k,\circ} \subseteq \{0,1\}$; 
  \vspace{0.3em}
  \State Compute non-conformity vectors using Eq.~(\ref{eq:nc_score}), for all $(x_i,y_i) \in \mathcal{D}_{\text{cal-base}}$. 
  \label{line:nc_score}
  \State Partition non-conformity space using $k$-means clustering on~$\mathcal{D}_{\text{cell}}$ to obtain cells $C_1, \ldots, C_K$. 
  \label{line:cell}
  \State Rank cells in ascending order of false-to-true ratio $\rho_m$ computed using Eq.~(\ref{eq:rho}): $C_{(1)}, C_{(2)}, \ldots, C_{(K)}$.
  \label{line:rank}
  \State Select top-ranked cells $\mathcal{C}_\alpha$ through Eq.~(\ref{eq:c_alpha}). 
  \label{line:top_cell}
  \State For new extraction: $\hat{y}^{\CONFORMAL}_{k,\circ} = \{y\in\{0,1\} : \mathbf{s}(y) \in \mathcal{C}_\alpha\}$.
  \label{line:inference}
\end{algorithmic}
\end{algorithm}
}

\vspace{0.3em}
\noindent\textbf{Coverage Guarantee.}
For any erroneous extraction (i.e., $y_{k,\circ} {=} 1$), %we have:
\begin{theorem}[Coverage Guarantee under Exchangeability]
Under the assumption that calibration and test examples are exchangeable, the conformal prediction set determined by \CONFORMAL satisfies:
\begin{equation*}
  \mathbb{P}\left( y_{k,\circ} \in \hat{y}^{\CONFORMAL}_{k,\circ} \right) \geq 1 - \alpha,
\end{equation*}
where $\hat{y}^{\CONFORMAL}_{k,\circ} = \left\{ y \in \{0,1\} \mid \mathbf{s}(y) \in \mathcal{C}_\alpha \right\}$.
\label{thm:scape_coverage}
\end{theorem}

\tech{
\begin{prf}
Assume the test example is exchangeable with the elements in $\mathcal{D}_{\text{re-cal}}^{\text{err}}$. 
Each example is assigned a non-conformity score $\mathbf{s}(1)$ (per Equation~(\ref{eq:nc_score})) and mapped into a cell $C_{(j)}$ among $K$ pre-defined, ranked cells. Let $R_i$ be the rank index of the lowest-ranked cell containing $\mathbf{s}_i(1)$, and $R_{\text{test}}$ be the corresponding rank for the test point.

The region $\mathcal{C}_\alpha$ is constructed by selecting the top $\eta^*$ cells such that at least $\lceil (1 - \alpha)(N_\text{re-cal} +1) \rceil$ calibration errors fall within them, as per Equation~(\ref{eq:c_alpha}), 
\begin{equation*}
\sum_{i=1}^{N_\text{re-cal}} \mathbf{1} \{ R_i \le \eta^* \} \ge \left\lceil (1 - \alpha)(N_\text{re-cal} +1) \right\rceil.
\end{equation*}
By exchangeability, $R_{\text{test}}$ is uniformly distributed among \{$R_1$, $\dots$, $R_N$, $R_{\text{test}}$\}, so 
\begin{equation*}
\mathbb{P}(R_{\text{test}} \le \eta^*) \ge 1 - \alpha.
\end{equation*}
Hence, $\mathbb{P}(\mathbf{s}_{\text{test}}(1) \in \mathcal{C}_\alpha) \ge 1 - \alpha$, which implies
\begin{equation*}
\mathbb{P}\left( y_{\text{test}} \in \hat{y}^{\CONFORMAL}_{\text{test}} \mid y_{\text{test}} = 1 \right) \ge 1 - \alpha.
\end{equation*}
\qed

\end{prf}
}

\vspace{0.3em}
\noindent\textbf{Set Size Optimality.} 
Theorem~\ref{thm:optimality} below establishes that \CONFORMAL achieves \emph{asymptotic optimality} in prediction set size under a mixture model assumption. This guarantees that, as the calibration data grows ($|\mathcal{D}_{\text{cal-base}}| \to \infty$), the method:
\begin{itemize}[itemsep=0pt, topsep=0pt, leftmargin=10pt]
  \item Minimizes the expected number of extractions flagged for human review ($\mathbb{E}[|\hat{y}^{\CONFORMAL}_{k,\circ}|]$), 
  \item While maintaining the desired error coverage ($1-\alpha$).
\end{itemize}
% The proof leverages the Neyman-Pearson lemma \cite{neyman1933ix,shafer2008tutorial} to show that ranking cells by false-to-true ratio $\rho_m$ (as Equation~(\ref{eq:rho})) is equivalent to optimizing the likelihood ratio $\Lambda(\mathbf{s})$ (see below). % UNCOMMENT IN TECHNICAL REPORT

The full proof, which leverages the Neyman-Pearson lemma \cite{neyman1933ix,shafer2008tutorial} to show that ranking cells by false-to-true ratio $\rho_m$ (as Equation~(\ref{eq:rho})) is equivalent to optimizing the likelihood ratio $\Lambda(\mathbf{s})$, is provided below. 

\begin{theorem}[Optimal Set Size of \CONFORMAL]
\label{thm:optimality}
  Assume that for each $c {\in} \{0,1\}$ the label-conditional densities $p(\mathbf{s}(c) \mid y{=}0)$ and $p(\mathbf{s}(c) \mid y{=}1)$ exist and are continuous ($y$ represents the true label). 
  Then, \CONFORMAL asymptotically minimizes the expected prediction set size $\mathbb{E}[|\hat{y}^{\CONFORMAL}_{k,\circ}|]$ subject to coverage $\geq 1-\alpha$. 
\end{theorem}

\tech{
\begin{prf}
We treat error detection as a binary hypothesis test $H_0: y {=} 0$ vs. $H_1: y {=} 1$, based on the observation $\mathbf{s}(0), \mathbf{s}(1) \in \mathbb{R}^{|\mathcal{L}|}$. For each extracted item, the prediction set $\hat{y}^{\CONFORMAL}_{k,\circ}$ is constructed by checking whether each candidate score vector $\mathbf{s}(y)$ lies in a selected region $\mathcal{C}_\alpha$ of the score space.

To construct $\mathcal{C}_\alpha$, we partition the score space into disjoint cells $\{C_m\}_{m=1}^K$ and compute the empirical false-to-true ratio $\rho_m$ in each cell using the definition in Equation~\ref{eq:rho}. 
As the calibration set size grows, $\rho_m$ converges to a population-level quantity:
\[
\lim_{|\mathcal{D}_{\text{re-cal}}| \to \infty} \rho_m \to 
\frac{
\int_{C_m} \left[ p(\mathbf{s}(0) \mid y{=}1) + p(\mathbf{s}(1) \mid y{=}0) \right] \, d\mathbf{s}
}{
\int_{C_m} \left[ p(\mathbf{s}(0) \mid y{=}0) + p(\mathbf{s}(1) \mid y{=}1) \right] \, d\mathbf{s}
}.
\]
Motivated by the Neyman-Pearson principle \cite{neyman1933ix,shafer2008tutorial}, we define the likelihood ratio for cell $C_m$ as:
\[
\Lambda(C_m) := 
\frac{
\int_{C_m} \left[ p(\mathbf{s}(0) \mid y=0) + p(\mathbf{s}(1) \mid y=1) \right] \, d\mathbf{s}
}{
\int_{C_m} \left[ p(\mathbf{s}(0) \mid y=1) + p(\mathbf{s}(1) \mid y=0) \right] \, d\mathbf{s}
}
= \frac{1}{\rho_m}.
\]

This likelihood ratio captures the total correct classification mass over misclassification mass in each cell, across both classes. Ranking cells by increasing $\rho_m$ is therefore asymptotically equivalent to ranking them by decreasing $\Lambda(C_m)$.

\CONFORMAL selects the smallest set of top-ranked cells $\mathcal{C}_\alpha$ such that the score vectors of at least $\lceil (1 - \alpha)(N_{\text{re-cal}} + 1) \rceil$ true errors are covered. The prediction set for a new point includes label $y$ if and only if $\mathbf{s}(y) \in \mathcal{C}_\alpha$.

The expected prediction set size is:
\[
\mathbb{E} \left[ \left| \hat{y}^{\CONFORMAL}_{k,\circ} \right| \right] = 
1 + \mathbb{P}\left( \mathbf{s}(1) \in \mathcal{C}_\alpha \mid y = 0 \right),
\]
which is minimized when $\mathcal{C}_\alpha$ contains cells with the highest $\Lambda$, i.e., lowest $\rho_m$, while satisfying the coverage constraint. Hence, under mild assumptions, \CONFORMAL asymptotically minimizes expected set size subject to valid coverage.
\qed
\end{prf}
}

\vspace{0.3em}
\noindent\textbf{Limitations.} 
Theorem~\ref{thm:optimality} establishes optimality under the assumption of a monotonic likelihood ratio between $\Lambda(\mathbf{s})$ and $\rho$. In practice, this requires the classifier outputs to be well-calibrated, which is challenging as we wish to keep |$\mathcal{D}_{\text{cal-base}}$| very small. Moreover, finite-sample effects may lead to marginal under-coverage when $|\mathcal{D}_{\text{cells}}|$ is small. 
Our empirical results demonstrate that extractions with high disagreement among layer-wise classifiers are more likely to be erroneous.
The raw non-conformity scores in \CONFORMAL (based on classifier probabilities) may not fully capture this disagreement. Metrics like $\kappa$---the number of layers disagreeing with the majority vote---provide an orthogonal signal that improves error detection. We thus extend \CONFORMAL below to capture such conflict. 

\subsection{\HYBRID: Hybrid Method}
\label{sec:hyb}

We now introduce \HYBRID, a unified method that incorporates the inter-layer conflict signal (\S\ref{sec:mv_cf}) into the conformal prediction framework (\S\ref{sec:scape}). The key idea is to augment the multi-dimensional non-conformity vector with a scaled conflict term, allowing the conformal predictor to respond not only to probabilistic uncertainty but also to internal disagreement among classifiers.

\vspace{0.3em}
\noindent\textbf{Conflict-Augmented Non-Conformity.}
% Recall from Equation~(\ref{eq:error_cf_kappa}) that the conflict score $\kappa$ is computed as the number of classifier layers whose binary prediction $\hat{y}_{k,\circ}^{(l)}$ disagree with the majority vote $\hat{y}_{k,\circ}^{\MAJORITY}$. 
The conflict score $\kappa$ used in CF (Equation~(\ref{eq:error_cf_kappa})) captures binary disagreement among classifiers, but tends to be, after probability thresholding per classifier, small or even zero; this happens even when the probability outputs of the classifiers vary significantly as they are ``smoothed'' by thresholding. Thus the value of $\kappa$ itself is not a very informative signal.
To address this, we replace $\kappa$ with a more granular, real-valued \textit{disagreement score} $\Delta$. 
Let $\pi^{(l)}$ denote the predicted probability of erroneous extraction from layer $l$, and let $\bar{\pi} = \frac{1}{|\mathcal{L}|} \sum_{l \in \mathcal{L}} \pi^{(l)}$ be the average probability across layers. 
We define the disagreement score as: 
$\Delta = \max_{l \in \mathcal{L}} |\pi^{(l)} - \bar{\pi}|$,
% \begin{equation} 
% \begin{aligned}
% \Delta = \max_{l \in \mathcal{L}} |\pi^{(l)} - \bar{\pi}|,
% \end{aligned} \end{equation}
which reflects the maximum deviation from consensus among the classifiers. 
We then embed it as an additional dimension into the non-conformity vector by defining a conflict-calibrated representation: 
\begin{equation} 
\begin{aligned}
\mathbf{s}_\lambda(c) = \Bigl[
(1{-}\lambda) \cdot s_1(c), \ldots, (1{-}\lambda) \cdot s_{|\mathcal{L}|}(c), \lambda \cdot {\Delta} \Bigr], 
\end{aligned} \label{eq:error_hyb_score} \end{equation}
where each $s_l(c)$ is the layer-wise non-conformity score defined in \S\ref{sec:scape}, and $\lambda {\in} [0,1]$ is a tunable parameter that adjusts the relative weight of the disagreement signal inside the conformal prediction pipeline. 
When $\lambda{=}0$, the method falls back to \CONFORMAL (\S\ref{sec:scape}); on the other hand, when $\lambda{=}1$, the method ignores all layer-wise non-conformity scores and relies exclusively on the disagreement score $\Delta$, effectively acting as a calibrated variant of conflict filtering (\S\ref{sec:mv_cf}).

\vspace{0.3em}
\noindent\textbf{Calibration and Prediction.}
Following the calibration protocol described in \S\ref{sec:scape}, we compute the conflict-augmented non-conformity vectors at the given $\lambda$-weighting for all examples in the calibration dataset and partition the non-conformity space into cells. 
We rank the cells through the conflict-augmented false-to-true ratio $\rho_m^\lambda$ and retain only the cells $\mathcal{C}_\alpha^\lambda$ that satisfy the conformal coverage condition at level $1 {-} \alpha$ (see Equation~(\ref{eq:c_alpha})). 
At test time, the conformal prediction set is defined as: 
\begin{equation*}
\hat{y}^{\text{Hyb}}_{k,\circ} = \left\{ y \in \{0,1\} \mid \mathbf{s}_\lambda(y) \in \mathcal{C}_\alpha^\lambda \right\},
\end{equation*}
As before, we treat the extraction as correct if $\hat{y}^{\text{Hyb}}_{k,\circ} {=} 0$, and trigger correction if the set includes 1 or both labels. 
It is easy to see that the conformal guarantee (Theorem~\ref{thm:scape_coverage}) still holds. This is because the augmented score $\mathbf{s}_\lambda(c)$ retains exchangeability, and the cell selection process is still thresholding a well-defined statistic.

\vspace{0.3em}
\noindent\textbf{Conflict-Aware Optimality.} 
\HYBRID extends the optimality guarantee of \CONFORMAL (Theorem~\ref{thm:optimality}) to conflict-augmented non-conformity scores $\mathbf{s}_\lambda$. Following the same principle as \CONFORMAL, it ranks cells in the score space by their empirical false-to-true ratio, which remains proportional to the inverse of the likelihood ratio. % measures the proportion of correct extractions misclassified as errors relative to true errors detected. 
By selecting the smallest set of cells $\mathcal{C}^\lambda_\alpha$ that satisfies the user-specified coverage constraint ($1{-}\alpha$), \HYBRID asymptotically minimizes the expected prediction set size. 

\tech{
\begin{theorem}[Optimality of \HYBRID]
\label{thm:hyb_optimality}
Let $\mathbf{s}_\lambda(c) \in \mathbb{R}^{|\mathcal{L}|+1}$ denote the conflict-augmented non-conformity score defined in Equation~(\ref{eq:error_hyb_score}). Suppose the class-conditional densities $p(\mathbf{s}_\lambda(c) \mid y)$ exist and are continuous for $y \in \{0,1\}$. Then, under the exchangeability and regularity assumptions of Theorem~\ref{thm:optimality}, the \HYBRID method asymptotically minimizes the expected prediction set size $\mathbb{E}[|\hat{y}^{\HYBRID}_{k,\circ}|]$ among all predictors satisfying class-conditional coverage $\geq 1 - \alpha$ for erroneous extractions ($y=1$).
\end{theorem}

\begin{prf}
\HYBRID augments the original non-conformity vector $\mathbf{s}(c) \in \mathbb{R}^{|\mathcal{L}|}$ with a real-valued conflict signal $\Delta \in \mathbb{R}$ to form the augmented vector $\mathbf{s}_\lambda(c) \in \mathbb{R}^{|\mathcal{L}| + 1}$, as defined in Equation~(\ref{eq:error_hyb_score}).

As in \CONFORMAL, the \HYBRID method partitions the augmented score space into disjoint cells $\{C_m^\lambda\}_{m=1}^{K}$ and ranks them using the empirical symmetric false-to-true ratio $\rho_m^\lambda$, defined analogously to Equation~(\ref{eq:rho}) but over the extended representation.

As the calibration size grows, this ratio converges to:
\[
\lim_{|\mathcal{D}_{\text{re-cal}}| \to \infty} \rho_m^\lambda \to 
\frac{
\int_{C_m^\lambda} \left[ p(\mathbf{s}_\lambda(0) \mid y=1) + p(\mathbf{s}_\lambda(1) \mid y=0) \right] \, d\mathbf{s}
}{
\int_{C_m^\lambda} \left[ p(\mathbf{s}_\lambda(0) \mid y=0) + p(\mathbf{s}_\lambda(1) \mid y=1) \right] \, d\mathbf{s}
}.
\]

We define the symmetric likelihood ratio for cell $C_m^\lambda$ as:
\[
\Lambda^\lambda(C_m^\lambda) :=
\frac{
\int_{C_m^\lambda} \left[ p(\mathbf{s}_\lambda(0) \mid y=0) + p(\mathbf{s}_\lambda(1) \mid y=1) \right] \, d\mathbf{s}
}{
\int_{C_m^\lambda} \left[ p(\mathbf{s}_\lambda(0) \mid y=1) + p(\mathbf{s}_\lambda(1) \mid y=0) \right] \, d\mathbf{s}
}
= \frac{1}{\rho_m^\lambda}.
\]

This generalizes the \CONFORMAL likelihood ratio to the conflict-augmented score space. Ranking cells by $\rho_m^\lambda$ thus corresponds asymptotically to ranking by decreasing $\Lambda^\lambda(C_m^\lambda)$.

\HYBRID selects the smallest set of top-ranked cells $\mathcal{C}^\lambda_\alpha$ to satisfy the desired coverage level. As in Theorem~\ref{thm:optimality}, the expected prediction set size is minimized by including only the most reliable cells (with highest $\Lambda^\lambda$ or lowest $\rho_m^\lambda$), while ensuring coverage.

Hence, \HYBRID asymptotically minimizes expected prediction set size under the coverage constraint, completing the proof.
\qed
\end{prf}
}

\vspace{0.3em}
\noindent\textbf{Advantage of \HYBRID.}
As long as the conflict score provides an additional discriminatory signal---specifically, if erroneous outputs tend to have higher conflict than correct ones---the hybrid method \HYBRID can yield smaller expected prediction sets (thus reducing human correction cost in expectation), while preserving the same coverage. 

\begin{theorem}[Optimality of \HYBRID over \CONFORMAL]
\label{thm:scape_hyb_advantage}
Assume that the conflict score provides an additional signal for distinguishing erroneous from correct extractions, i.e., erroneous examples ($y{=}1$) are more likely to have a higher conflict score than correct ones. 
Then, under the same $(1 {-} \alpha)$ coverage constraint, \HYBRID produces a prediction set with equal or smaller expected size compared to \CONFORMAL:
\begin{equation*}
  \mathbb{E}\left[ \left| \hat{y}^{\textsc{Hyb}}_{k,\circ} \right| \right] \leq \mathbb{E}\left[ \left| \hat{y}^{\CONFORMAL}_{k,\circ} \right| \right] .
\end{equation*}
\end{theorem}

\tech{
\begin{assumption}[Disagreement Signal Consistency]
\label{assumption:delta}
The disagreement score $\Delta$ is more likely to take high values under erroneous extractions (true label $y{=}1$) than under correct ones. In other words, the conditional density of $\Delta$ is higher under $y{=}1$ than under $y{=}0$:
\begin{equation*}
\frac{p(\Delta \mid y=1)}{p(\Delta \mid y=0)} \geq 1.
\end{equation*}
\end{assumption}

A detailed formal proof of Theorem~\ref{thm:scape_hyb_advantage} is provided below. 
\begin{prf}

We begin with the following definitions. 
Let $d{=}|\mathcal{L}|$, and define the original non-conformity vector space used by \CONFORMAL as $\mathcal{V} {=} \mathbb{R}^d$. Define the augmented space used by \HYBRID as $\mathcal{V}' {=} \mathbb{R}^{d+1}$.
Define a projection operator $\delta: \mathcal{V}' {\rightarrow} \mathcal{V}$ as: 
\begin{equation*}
\delta\left( \left[ s_1,\dots,s_d,s_{d+1} \right] \right) = \left[s_1,\dots,s_d \right].
\end{equation*}

\noindent
1. Neyman-Pearson Optimality (\CONFORMAL).
Based on the Neyman-Pearson lemma \cite{neyman1933ix}, the optimal acceptance region for \CONFORMAL in $\mathcal{V}$ is given by: 
\begin{equation*} \begin{aligned}
  R_{\CONFORMAL} = \left\{\mathbf{s}\in\mathcal{V}:\Lambda(\mathbf{s}) > t_{\CONFORMAL} \right\},
\end{aligned} \end{equation*}
where the likelihood ratio is $\Lambda(\mathbf{s}) = p(\mathbf{s}(1) | y{=}1) / p(\mathbf{s}(1) | y{=}0)$. 
The threshold $t_{\CONFORMAL}$ is chosen to satisfy:
\begin{equation*}
\Pr_{y=1}[\mathbf{s}(1) \in R_{\CONFORMAL}] = 1 - \alpha.
\end{equation*}

\noindent
2. Extended Acceptance Region (\CONFORMAL).
Define the inverse projection of $R_{\CONFORMAL}$ in the extended space $\mathcal{V}'$:
\begin{equation*}
  \widetilde{R}_{\CONFORMAL}
    = \delta^{-1}\left(R_{\CONFORMAL}\right)
    = \left\{ \mathbf{s}' \in \mathcal{V}':\delta(\mathbf{s}')\in R_{\CONFORMAL} \right\}, 
\end{equation*}
This set does not differentiate values in the added $(d+1)$-th dimension, thus preserving coverage and false-positive rate:
\begin{equation*} \begin{aligned}
  &\Pr_{y=1}[\mathbf{s}'(1) \in \widetilde{R}_{\CONFORMAL}] = \Pr_{y=1}[\mathbf{s}(1) \in R_{\CONFORMAL}] = 1-\alpha, \\
  &\Pr_{y=0}[\mathbf{s}'(0) \in \widetilde{R}_{\CONFORMAL}] = \Pr_{y=0}[\mathbf{s}(0) \in R_{\CONFORMAL}].
\end{aligned} \end{equation*}

\noindent
3. Likelihood Ratio Advantage (\HYBRID).
In the extended space $\mathcal{V}'$, under the Assumption~\ref{assumption:delta}, the likelihood ratio function $\Lambda_\lambda$ of \HYBRID satisfies:
\begin{equation*} 
\begin{aligned}
  \Lambda_\lambda(\mathbf{s}_\lambda) 
  &= \frac{p(\mathbf{s}_\lambda(1) \mid y{=}1) + p(\mathbf{s}_\lambda(0) \mid y{=}0)}{p(\mathbf{s}_\lambda(1) \mid y{=}0) + p(\mathbf{s}_\lambda(0) \mid y{=}1)} \\
  &= \frac{
    p(\mathbf{s}(1) \mid y{=}1) \cdot p(\Delta \mid y{=}1)
    + p(\mathbf{s}(0) \mid y{=}0) \cdot p(\Delta \mid y{=}0)
  }{
    p(\mathbf{s}(1) \mid y{=}0) \cdot p(\Delta \mid y{=}0)
    + p(\mathbf{s}(0) \mid y{=}1) \cdot p(\Delta \mid y{=}1)
  } \\
  &\ge \frac{p(\mathbf{s}(1) \mid y{=}1) + p(\mathbf{s}(0) \mid y{=}0)}{p(\mathbf{s}(1) \mid y{=}0) + p(\mathbf{s}(0) \mid y{=}1)}
  = \Lambda(\delta(\mathbf{s}_\lambda)),
\end{aligned}
\end{equation*}
Thus, using the same threshold $t_{\CONFORMAL}$, we have:
\begin{equation*}
  \Pr_{y=1}\left[ \Lambda_\lambda(\mathbf{s}_\lambda) > t_{\CONFORMAL} \right]
  \ge
  \Pr_{y=1}\left[ \Lambda(\delta(\mathbf{s}_\lambda)) > t_{\CONFORMAL} \right]
  = 1 - \alpha.
\end{equation*}

\noindent
4. Coverage Matching (\HYBRID).
To achieve exact $(1{-}\alpha)$ coverage, we select $t_{\textsc{Hyb}} \ge t_{\CONFORMAL}$ such that: 
\begin{equation*}
  \Pr_{y=1}\bigl[ \Lambda_\lambda(\mathbf{s}_\lambda) > t_{\textsc{Hyb}} \bigr] = 1-\alpha.
\end{equation*}
Since increasing the threshold reduces the coverage set and the coverage rate decreases monotonically, there must exist such a $t_{\textsc{Hyb}}$. 

\noindent
5. Acceptance Region Comparison.
Define the acceptance region for \HYBRID with threshold $t_{\textsc{Hyb}}$:
\begin{equation*}
  R_{\textsc{Hyb}}
    = \{\mathbf{s}'\in\mathcal S':\Lambda_\lambda(\mathbf{s}') > t_{\textsc{Hyb}}\}.
\end{equation*}
Given $t_{\textsc{Hyb}} \ge t_{\CONFORMAL}$ and $\Lambda_\lambda \ge \Lambda \circ \delta$, for any $\mathbf{s}'\in \mathcal{V}'$: 
\begin{equation*}
  \Lambda_\lambda(\mathbf{s}') > t_{\textsc{Hyb}}
  \ \Longrightarrow\ 
  \Lambda_\lambda(\mathbf{s}') > t_{\CONFORMAL}
  \ \Longrightarrow\ 
  \Lambda(\delta(\mathbf{s}')) > t_{\CONFORMAL},
\end{equation*}
Thus, 
\begin{equation*}
  R_{\textsc{Hyb}} \subseteq \{\mathbf{s}': \Lambda(\delta(\mathbf{s}')) > t_{\CONFORMAL}\}
  = \delta^{-1}(R_{\CONFORMAL}).
\end{equation*}

\noindent
6. Expected Set Size Advantage.
The original \CONFORMAL method defines its acceptance region $R_{\CONFORMAL} \subseteq \mathcal{V}$ in the original non-conformity space. Its corresponding inverse projection (or cylindrical lift) in the extended space is:
\begin{equation*}
  \delta^{-1}(R_{\CONFORMAL}) = \left\{ \mathbf{s}' \in \mathcal{V}': \delta(\mathbf{s}') \in R_{\CONFORMAL} \right\},
\end{equation*}
which retains the same false positive rate as $R_{\CONFORMAL}$, since the additional coordinate has no effect on the classifier outputs:
\begin{equation*}
  \Pr_{y=0}[\mathbf{s}' \in \delta^{-1}(R_{\CONFORMAL})] = \Pr_{y=0}[\mathbf{s} \in R_{\CONFORMAL}].
\end{equation*}
In contrast, the acceptance region $R_{\textsc{Hyb}}$ used by \HYBRID is a strict subset of this lifted region, as established in Step 5:
\begin{equation*}
  R_{\textsc{Hyb}} \subseteq \delta^{-1}(R_{\CONFORMAL}).
\end{equation*}
Therefore, it must hold that the false positive rate of \HYBRID is at most that of \CONFORMAL:
\begin{equation*}
  \Pr_{y=0}[\mathbf{s}'\in R_{\textsc{Hyb}}]\le\Pr_{y=0}[\mathbf{s}\in R_{\CONFORMAL}],
\end{equation*}
which implies directly:
\begin{equation*}
  \mathbb{E}\left[ \left| \hat{y}^{\textsc{Hyb}}_{k,\circ} \right| \right] \leq \mathbb{E}\left[ \left| \hat{y}^{\CONFORMAL}_{k,\circ} \right| \right].
\end{equation*}
\qed

\end{prf}
}

Our experimental results (\S\ref{sec:exp}) empirically corroborate this theoretical advantage. 

Besides, \HYBRID can be viewed as a \emph{soft} compatible generalization of conflict filtering: instead of enforcing hard thresholds, the conflict score is smoothly embedded into the non-conformity space and calibrated within the conformal framework. In this way, the hybrid method preserves a key advantage of \CONFLICT---its ability to improve recall at relatively low cost---while avoiding brittle thresholds and retaining the formal coverage guarantees of \CONFORMAL. The continuous weighting of conflict also gives practitioners finer control over the accuracy-cost trade-off.

\vspace{0.3em}
\noindent\textbf{Summary.} 
The parameter $\lambda$ controls the relative influence of conflict: higher values give more weight to conflict, approaching \CONFLICT behavior as $\lambda \to 1$; lower values emphasize the original conformal score. 
Low values in $\lambda$ prioritize probabilistic uncertainty and are better for well-calibrated classifiers (when |$\mathcal{D}_{\text{cls}}$| is large). High values of $\lambda$ prioritize conflict, which is better when layer disagreements correlate with errors and as we empirically demonstrate in \S\ref{sec:exp_population} when |$\mathcal{D}_{\text{cls}}$| is small. The optimal setting can be selected via grid search on a validation set. In practice, intermediate values (e.g., $\lambda {=} 0.5$) often yield a good trade-off between error detection recall and extra correction cost, as demonstrated in \S\ref{sec:exp_population}.

\section{Iterative Schema Discovery}
\label{sec:schema}

The goal of this stage is to derive a relational schema $S_D^q$ over a collection of document chunks $D$ that contains exactly the entities (attributes, such as names and locations) and relationships required to answer the query $q$. 
Schema discovery occurs dynamically in two phases. Phase I induces a general (query-agnostic) schema $S_D^\text{gen}$, that captures all salient attributes and relationships present in the documents, independently of any specific query. Phase II then adapts this into a query-specific schema $S_D^q$ tailored explicitly to the requirements of query $q$.

% Phase I
\vspace{0.3em}
\noindent\textbf{Phase I: General Schema Discovery.}
We treat schema discovery as an iterative process of reading and abstraction. It begins with an empty schema and processes the document chunks in a sequential, one-pass manner. At each step, it incrementally revises the current schema state. As new information becomes available, each step may refine earlier decisions by revising the previously constructed schema state. The final schema is the result of this sequence of incremental updates across all chunks. We next describe the structure of the schema state as maintained during this process, i.e., how the algorithm represents and updates it at each step.

The schema state is organized as a collection of relational tables. Each table corresponds to either an entity type (e.g., \smallcode{Person}, \smallcode{Hospital}) or a relationship (e.g., \smallcode{Admission}, \smallcode{Treatment}). 
To help the algorithm maintain and use the schema state effectively, each table is annotated with the following information.
\begin{itemize}[itemsep=0pt, topsep=0pt, leftmargin=10pt]
  \item A canonical table name and a concise natural language description, both generated by the LLM based on the semantics of relevant document chunks and prior schema context;
  \item A small set of example document chunks for each table, selected by the LLM to motivate the creation of the table and provide grounding for its semantics; 
  \item A list of attributes for each table, each annotated with a name and a usage-based explanation derived by the LLM from the context in which it appeared. 
\end{itemize}

Schema updates are performed by a prompt-based function $\mathcal{S}_G$, which uses a fixed prompt template to invoke an LLM\footnote{\tech{Details of the prompt design and its implementation are available in Appendix~\ref{apx:prompt_general}.}}. Given the current schema state $S_{k-1}^\text{gen}$ and the $k$-th document chunk $d_k$, the algorithm computes the next state as:
\begin{equation}
  S_k^\text{gen} = \mathcal{S}_G\left( S_{k-1}^\text{gen}, d_k \right), \quad \text{for } k = 1..n
\end{equation}
Updates may involve introducing new tables, adding attributes to existing tables, or refining existing descriptions. Crucially, the current schema state remains in memory at every step, enabling the function $\mathcal{S}_G$ to leverage prior schema structure and annotations to extract additional structure information from each new chunk.

As the process continues, earlier schema elements that were ambiguous or incomplete may be clarified by later chunks, supporting limited self-correction without retroactive reconstruction. We empirically validate the effectiveness of our method in \S\ref{sec:exp_schema}.
After all chunks are processed, the final schema $S_n^\text{gen}$ constitutes $S_D^\text{gen}$. It serves as a comprehensive, query-agnostic abstraction over the document collection and provides the structural basis for the next stage of query-specific schema adaptation.

% Phase II
\vspace{0.3em}
\noindent\textbf{Phase II: Query-Specific Schema Discovery.} 
In the second phase, the goal is to transform the general schema $S_D^\text{gen}$ into a query-specific schema $S_D^q$ that contains only the tables and attributes necessary to answer the input query $q$. The process is again iterative, following the same schema structure and update pattern as in Phase I. Schema updates are now performed by a different prompt-based function $\mathcal{S}_Q$, which also uses a fixed LLM prompt\footnote{\tech{Details of the prompt design and its implementation are available in Appendix~\ref{apx:prompt_query}.}}, but incorporates the query $q$ as an additional input to guide the refinement. 
\begin{equation}
S_k^q = \mathcal{S}_Q\left( S_{k-1}^q, d_k, q, S_D^\text{gen} \right), \quad \text{for } k = 1..n
\end{equation}
In this phase, $\mathcal{S}_Q$ selectively removes irrelevant schema elements, adds previously overlooked attributes, if any, and restructures attributes to precisely match the intent of query $q$. Such refinement includes accommodating explicit query constraints (e.g., filters, group-by keys) and implicit query requirements (e.g., join paths, derived attributes). Similar to Phase I, schema decisions can be iteratively revised if errors are introduced in earlier steps.

The result is a minimal and query-complete schema $S_D^q$, essential for accurate data population and effective query execution. Without this targeted refinement, critical attributes might be omitted or extraneous attributes retained, impairing query effectiveness. 
We empirically demonstrate in \S\ref{sec:exp_schema} that omitting this step leads to lower schema completeness and reduced query accuracy. 

\vspace{0.3em}
\noindent\textbf{Repair Step.} 
As an optional enhancement, a \textit{fallback repair} step can be applied after Phase II, using a powerful LLM (e.g., GPT-5~\cite{openai_gpt5}) to verify whether the extracted schema $S_D^q$ suffices to answer the query $q$; if not, the system re-invokes schema discovery with additional iterations to repair and complete the schema. 
We empirically show in \S\ref{sec:exp_schema} that this repair mechanism discovers any attributes, achieving perfect attribute-level recall across all datasets.

\vspace{0.3em}
\noindent\textbf{Performance Considerations.} 
Our schema discovery pipeline performs two sequential passes over the document collection: 
Phase I builds a general schema from scratch, and Phase II refines it to match the specific query intent. 
While techniques such as sampling, chunk clustering, or selective analysis could significantly reduce computational cost, they are orthogonal to the focus of this work.  
We aim to understand whether one can indeed design an \textit{accurate} schema discovery and data extraction strategy for a specific query without hard computational considerations (e.g., token usage \cite{satriani2025logical, liu2025palimpzest}). 
Answering this question first, instigates future research directions or engineering optimizations to derive performance efficiency.
A detailed exploration of efficiency-oriented enhancements is left to future work and falls outside the scope of this paper.

\section{Experimental Evaluation}
\label{sec:exp}

\subsection{Experimental Setup}

\subsubsection{Datasets.}
~
We evaluate \MAINNAME on {\em five datasets} that simulate realistic information extraction and analytical query scenarios over unstructured or weakly structured document collections.
% 加一句话，一些结构化或者半结构化的 ，相对于LLM 提取来说过于简单的 数据集入 SWDE 和 %KarmaBench 没有被考虑 
Table~\ref{tab:exp_datasets} summarizes these datasets, detailing the number of queries per set and the average number of expected output table entries per query.
The first two datasets, \SPIDER and \BIRD, are derived from the Spider~\cite{dataset_spider} and Bird~\cite{dataset_bird} benchmarks. Using their original schemas and tabular data, following \cite{maekawa2025holistic} we convert tabular rows into natural language document chunks using a state-of-the-art LLM (GPT-5~\cite{openai_gpt5}).
This ensures that the model used in \MAINNAME (Qwen3-30B-A3B \cite{qwen3technicalreport}) has not seen these documents during training, thereby mitigating data leakage concerns.
For each benchmark query, we know the precise result (ground truth) to evaluate correctness.
\MAINNAME is evaluated on the original natural language queries from Spider and Bird datasets, as well as on newly introduced queries that involve multi-table joins, aggregation, and multi-hop reasoning.
In our evaluation, we do not account for natural language to SQL translation; instead, we provide the correct SQL query to execute on the extracted tables. We do this to isolate our evaluation from text-to-SQL translation errors; naturally any state-of-the-art text-to-SQL technique can be adopted.
While generating the datasets, we randomly shuffle the chunks to eliminate any semantic correlations.
We also introduce varying degrees of information density in the document collection (ratio of relevant vs irrelevant chunks to the query) in \S\ref{sec:exp_additional}.
\tech{The prompts used to generate the documents, due to space constraints, are provided in Appendix~\ref{apx:prompt_dataset}.}
The dataset \Galois is sourced from the \KAGGLE and \BBC datasets as described in \cite{satriani2025logical}.
The dataset \FDA is sourced from FDA 510(k) regulatory filings \cite{wu2021medical, arora2023language}, which are long-form and heterogeneous, containing narrative summaries, tabular sections, and metadata. 
The dataset \CUAD is a legal-contract benchmark \cite{shankar2024docetl}, whose lengthy documents make single-pass LLM processing infeasible due to context limitations \cite{shankar2024docetl}.
% chunk???
For these datasets, we use both the original benchmark queries (typically involving a limited set of documents and attributes) and additional new queries proposed in this paper. 
The additional queries are designed to include more attributes and to incorporate cross-document aggregation as well as multi-hop reasoning\footnote{We plan to make available all artifacts associated with this paper.}.

\begin{table}[t]
  \centering \small
  \captionsetup{font=small, skip=4pt}
  \caption{
    Evaluation Datasets Used for Assessing \MAINNAME. 
  }
  \label{tab:exp_datasets}
  \begin{tabular}{p{1.6cm} l c c}
  \toprule
  \textbf{Dataset} & \textbf{Dataset Source} & \textbf{\# Queries} & \textbf{Avg. Result Rows} \\
  \midrule
  \SPIDER & Spider \cite{dataset_spider} & 86 & 1733 \\
  \BIRD & Bird \cite{dataset_bird} & 36 & 2435 \\
  \Galois & Galois \cite{satriani2025logical} & 10 & 497 \\
  \FDA & FDA 510(k) \cite{wu2021medical} & 6 & 100 \\
  \CUAD & CUAD \cite{shankar2024docetl} & 15 & 501 \\
  \bottomrule
  \end{tabular}
\end{table}

\subsubsection{Measurements.}
~
We evaluate \MAINNAME in two stages: schema discovery (ISD) and data population (TDP). 
For data population, we evaluate the accuracy of the extracted tabular data by comparing it with the ground truth at the \textit{cell} (attribute value) level. 
Specifically, each ground-truth cell (i.e., each populated value in the ground-truth table) is checked against the extracted result. 
We then compute the following accuracy metric \cite{shankar2024docetl}: 
\begin{equation}
  \textit{ACC}_\textit{pop} = 1 - \frac{\text{\# missing cells} + \text{\# incorrect cells}}{\text{\# ground-truth cells}}
  \label{eq:exp_acc_p}
\end{equation}
Here, \textit{missing} cells are those that should have been extracted but were not, and \textit{incorrect} cells are those that were extracted but contain erroneous values.

\CONFORMAL and \HYBRID, identify potentially erroneous cells and use additional review steps (e.g., human inspection) to validate and fix extraction errors. 
If correctly extracted cells are flagged for inspection, this results in unnecessary correction efforts. 
To quantify this, we compute the false positive rate:
\begin{equation}
  \textit{FPR}_\textit{pop} = \frac{\textit{FP}}{\textit{FP} + \textit{TN}},
  \label{eq:exp_fpr_p}
\end{equation}
where $\textit{FP}$ (false positives) denotes correctly extracted cells flagged for inspection and $\textit{TN}$ (true negatives) denotes correctly extracted cells not flagged. 
For a fixed $\textit{ACC}_\textit{pop}$, a higher $\textit{FPR}_\textit{pop}$ implies additional wasted effort inspecting accurate extractions, while a lower $\textit{FPR}_\textit{pop}$ reflects a more efficient process, focusing inspections primarily on erroneous extractions.
Thus, $\textit{FPR}_\textit{pop}$ quantifies the inefficiency (unnecessary inspections) in the extraction pipeline.

For schema discovery, we first assess whether the discovered schema for each query is sufficient to answer the query, resulting in an accuracy metric denoted as $\textit{ACC}_\textit{sch}$. 
Additionally, by comparing the discovered schema with the ground-truth schema in terms of their attributes, we compute attribute-level recall ($\textit{Rec}_\textit{sch}^\textit{attr}$) and precision ($\textit{Pre}_\textit{sch}^\textit{attr}$), which measure the completeness and redundancy of the discovered schema.

\subsubsection{Experimental Settings.}
~
\MAINNAME is implemented in Python. For schema discovery, it employs GPT-5 \cite{openai_gpt5}, a state-of-the-art LLM. For data population, it employs Qwen3-30B-A3B \cite{qwen3technicalreport}, a leading open-source model at the 30B scale, deployed on an A100 (80GB) GPU. 
Unless otherwise specified, all experiments use the following default settings: the conflict weight parameter is set to $\lambda {=} 0.5$ (see Equation~(\ref{eq:error_hyb_score})); the classifiers are trained with 50 entries and calibrated on 150 entries randomly sampled per query.

\begin{table}[t]
  \centering \small
  \captionsetup{font=small, skip=4pt}
  \caption{
    Summary of Data Population Accuracy ($\textit{ACC}_\textit{pop}$). 
    \textnormal{
      All values are averages over queries in Tb.~\ref{tab:exp_datasets}. 
      $\textit{ACC}_\textit{pop}$ is defined in Eq.~(\ref{eq:exp_acc_p}). 
    } 
  }
  \label{tab:exp_population_accuracy}
  \begin{tabular}{p{2.5cm} c c c c c}
  \toprule
  \textbf{Method} & \SPIDER & \BIRD & \Galois & \FDA & \CUAD \\
  \midrule
  EVAPORATE \cite{arora2023language} & - & - & 0.475 & 0.516 & 0.209 \\
  Palimpzest \cite{liu2025palimpzest} & - & - & 0.867 & 0.924 & 0.613 \\
  \MAINNAME (No Correction) & 0.938 & 0.949 & 0.873 & 0.965 & 0.661 \\
  \MAINNAME (\CONFORMAL) & 0.991 & 0.992 & 0.989 & 0.988 & 0.724 \\
  \MAINNAME (\HYBRID) & \textbf{0.993} & \textbf{0.994} & \textbf{0.989} & \textbf{0.990} & \textbf{0.983} \\
  \bottomrule
  \end{tabular}
\end{table}

\begin{table}[t]
  \centering \small
  \captionsetup{font=small, skip=4pt}
  \caption{
    Correction Overhead in Data Population. 
    \textnormal{
      All values are averages over queries in Tb.~\ref{tab:exp_datasets}. 
      $\textit{FPR}_\textit{pop}$ quantifies the cost of unnecessary corrections, defined in Eq.~(\ref{eq:exp_fpr_p}). 
    }
  }
  \label{tab:exp_population_overhead}
  \begin{tabular}{p{2.5cm} c c c c c}
  \toprule
  \textbf{Method} & \SPIDER & \BIRD & \Galois & \FDA & \CUAD \\
  \midrule
  \MAINNAME (\CONFORMAL) & 0.063 & 0.054 & 0.044 & 0.051 & 0.114 \\
  \MAINNAME (\HYBRID)    & \textbf{0.038} & \textbf{0.039} & \textbf{0.027} & \textbf{0.032} & \textbf{0.072} \\
  \bottomrule
  \end{tabular}
\end{table}

\subsection{Results of Data Population and Correction}
\label{sec:exp_population}

\subsubsection{Accuracy and Correction Cost.}
\label{sec:exp_population_acc}
~
Table~\ref{tab:exp_population_accuracy} reports the data population accuracy ($\textit{ACC}_\textit{pop}$, defined in Equation~\ref{eq:exp_acc_p}) under different configurations proposed in this paper: (i) \MAINNAME without Correction, which executes without attempting to detect errors (outputs the extraction directly from the LLM), (ii) \CONFORMAL, and (iii) \HYBRID, which integrates the proposed algorithms with default parameters ($\alpha{=}0.15$), as well as prior baselines, EVAPORATE \cite{arora2023language} and Palimpzest \cite{liu2025palimpchat, liu2025palimpzest}\footnote{
  \cite{satriani2025logical} in their recent evaluation, demonstrated that Palimpzest has superior accuracy over other prior baselines. For this reason we compare with Palimpzest, without including the baselines that \cite{satriani2025logical} demonstrated inferior to Palimpzest.  We do this as ccuracy is the focus of our work. We run experiments with the same configuration parameters as in \cite{satriani2025logical}.
}.
All metrics are averaged over all queries in Table~\ref{tab:exp_datasets}.

Without correction, \MAINNAME achieves reasonably high accuracy (e.g., 0.938 on \SPIDER and 0.949 on \BIRD), but still produces a substantial number of error cells (e.g., 516 remaining error cells on \SPIDER and 491 on \BIRD). {\em Since EVAPORATE and Palimpzest are not designed to handle schemas involving multiple tables, we present their evaluation only on \Galois, \FDA, and \CUAD datasets (which do not involve multiple tables)}. Using the same base model (Qwen3-30B-A3B \cite{qwen3technicalreport}), EVAPORATE performs considerably worse than other approaches, while Palimpzest achieves accuracy comparable (though slightly lower) than \MAINNAME {\em without correction} (e.g., 0.867 vs. 0.873 on \Galois). 

By contrast, applying our correction algorithms yields substantial improvements. With $\alpha{=}0.15$, both \CONFORMAL and \HYBRID raise accuracy to {\bf above or near 0.99} on \SPIDER, \BIRD, \Galois, and \FDA, with \HYBRID consistently achieving the highest accuracy across all datasets. 

The CUAD dataset poses unique challenges due to its long-document nature: a single legal contract (corresponding to one row in the ground-truth table) can be nearly 100,000 tokens, far exceeding the context window of Qwen3-30B-A3B \cite{qwen3technicalreport}. 
To address this, we adapt a map-reduce style document chunking strategy inspired by DocETL \cite{shankar2024docetl}. Each long contract is divided into smaller chunks (e.g., sections or pages) that can fit within the LLM's context window. Instead of assuming that one document chunk corresponds to one complete table row (as in the default setting), we allow multiple chunks (from the same legal contract) to collectively provide the attribute values required for a single row. 
% For instance, in CUAD a contract's ``termination clause'' and ``payment obligation'' may appear in different sections of the same document; both must be processed and merged to populate the row representing that contract. 
Attribute values are first extracted independently from each chunk, and then consolidated into a complete row.
Combined with our correction algorithm, this chunk-merge (map-reduce) pipeline enables \HYBRID to reach 0.983 accuracy on CUAD---substantially higher than \CONFORMAL (0.724, without chunk merge) and Palimpzest (0.583).

Table~\ref{tab:exp_population_overhead} reports on the corresponding correction overhead in terms of false positive rate ($\textit{FPR}_\textit{pop}$). 
Here, \HYBRID consistently maintains lower false positive rates than \CONFORMAL across all datasets (e.g., 0.038 vs. 0.063 on \SPIDER and 0.039 vs. 0.054 on \BIRD), indicating fewer correctly extracted cells are unnecessarily flagged for review. The overhead remains modest overall, particularly on large-scale benchmarks where \HYBRID achieves both higher accuracy and lower correction cost. 
These results demonstrate the effectiveness of our correction module in improving extraction quality while incurring minimal verification overhead. 

\begin{figure}[t]
  \centering
  \captionsetup{font=small, skip=0pt}
  \includegraphics[width=0.48\textwidth]{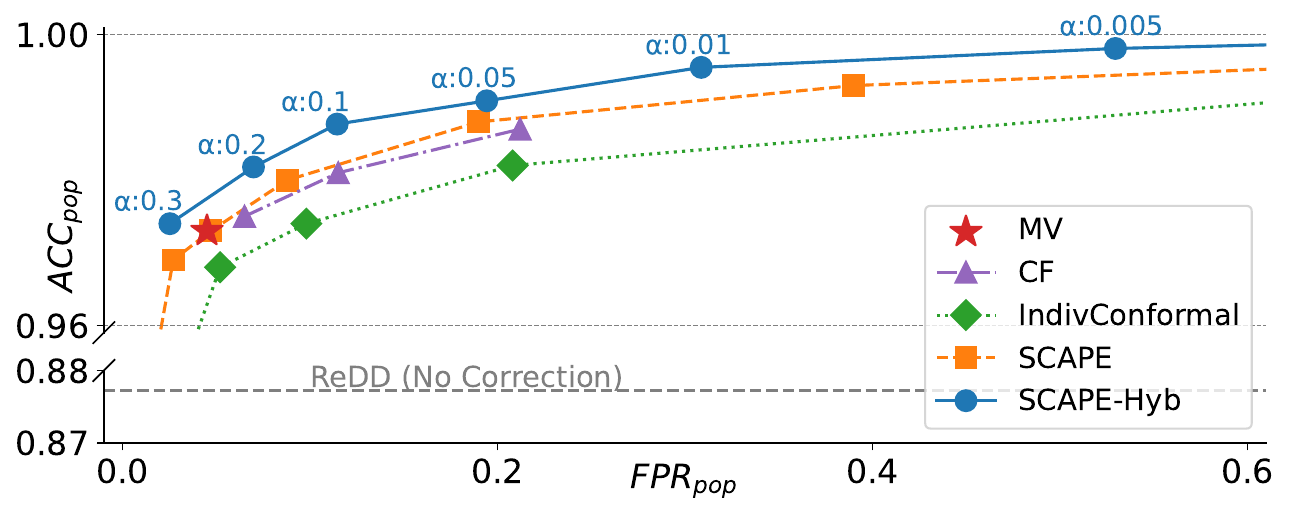}
  \caption{
    Trade-off between data population accuracy ($\textit{ACC}_\textit{pop}$) and correction cost measured by the false positive rate ($\textit{FPR}_\textit{pop}$). 
    \textnormal{
      For the \HYBRID curve, labels above each point indicate the corresponding $\alpha$ value used to produce that accuracy–cost trade-off.
    }
  }
  \label{fig:exp_acc_cost_all}
\end{figure}

\begin{figure}[t]
  \centering
  \captionsetup{font=small, skip=0pt}
  \centering
  \includegraphics[width=0.48\textwidth]{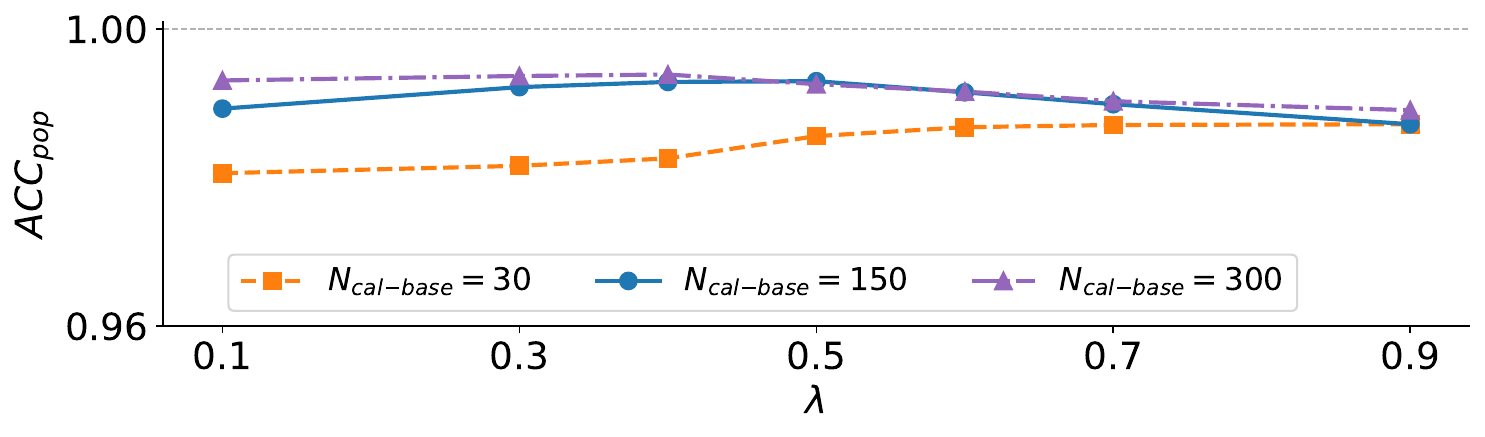}
  \caption{
    Data population accuracy $\textit{ACC}_\textit{pop}$ for \HYBRID with different calibration dataset size $N_\text{cal-base}$ on dataset \SPIDER, varying conflict weight $\lambda$, under $\textit{FPR}_\textit{pop}{=}0.2$. 
  }
  \label{fig:exp_diff_lambda}
\end{figure}

\subsubsection{Ablation: Accuracy-Cost Tradeoff.}
\label{sec:exp_population_ablation}
~
We conduct ablation experiments to compare \HYBRID against several error detection methods proposed in this paper: MV (\S\ref{sec:mv_cf}), CF (\S\ref{sec:mv_cf}), and \CONFORMAL. We also evaluate an alternative method that calibrates each base classifier independently using conformal prediction \cite{angelopoulos2022gentleintroductionconformalprediction}, followed by majority voting over the calibrated outputs (denoted as \textit{IndivConformal}). 
The results are presented in Figure~\ref{fig:exp_acc_cost_all}, which plots data population accuracy ($\textit{ACC}_\textit{pop}$) against unnecessary correction cost measured by the false positive rate ($\textit{FPR}_\textit{pop}$) averaged over all datasets. 
The gray dashed line marks the accuracy of \MAINNAME without correction. 
All methods yield notable improvements over the baseline. 
The MV approach, lacks tunable parameters and provides only a fixed trade-off point. The curve for IndivConformal consistently lies below others, indicating weaker performance. CF supports limited tuning but remains inferior in accuracy---its curve consistently lies below \CONFORMAL, indicating higher correction cost for the same level of accuracy. 
In contrast, \CONFORMAL and \HYBRID offer significantly better accuracy-correction cost trade-offs. \HYBRID consistently outperforms all baselines. It achieves the highest data population accuracy while incurring the lowest rate of unnecessary corrections, demonstrating the effectiveness of combining \CONFORMAL with conflict information. 

\subsubsection{Effect of Coverage Threshold $\alpha$.}
\label{sec:exp_population_alpha}
~
The results in Figure~\ref{fig:exp_acc_cost_all}, averaged across all datasets, reveal a key trend: \HYBRID achieves high initial accuracy (${\geq}0.974$ across all datasets) even at large $\alpha$, with more accuracy gains as $\alpha$ decreases. For small $\alpha$, accuracy can reach 1, but at the cost of a higher false positive rate ($\textit{FPR}_\textit{pop}$). 

Using the \SPIDER dataset as an example (avg. 1,733 rows per query), setting $\alpha{=}0.5$ yields 0.947 accuracy with 4.9\% of the data reviewed (including 8 false positives). Reducing the threshold to $\alpha{=}0.3$ improves accuracy to 0.975, with 9.2\% of rows reviewed (37 false positives). At $\alpha{=}0.01$, accuracy further increases to 0.998, but 31\% of the data must be reviewed (608 false positives).
As $\alpha$ decreases, the proposed framework becomes more conservative---fewer predictions are accepted as confident, and more data are flagged for review. This generally improves accuracy, as more uncertain predictions are reviewed and corrected. However, it also increases false positives, reflected in higher $\textit{FPR}_\textit{pop}$. Conversely, larger $\alpha$ values yield cheaper but relatively less accurate results. 

The results suggest that minimal user effort (relative to output size) suffices for high accuracy (${>}0.97$), while perfect accuracy (1.0) demands significantly more verification effort. This finding opens several research directions, as discussed in \S\ref{sec:con}.

\subsubsection{Effect of Conflict Weight $\lambda$.}
\label{sec:exp_population_lambda}
~
Figure~\ref{fig:exp_diff_lambda} illustrates the impact of the conflict weight ($\lambda$) on data population accuracy ($\textit{ACC}_\textit{pop}$) for \HYBRID on dataset \SPIDER (under $\textit{FPR}_\textit{pop}{=}0.2$). 
According to the blue curve under default setting calibration dataset size $N_\text{cal-base}{=}150$ the results reveal a clear trend: accuracy peaks at $\lambda {\approx} 0.5$ and declines when $\lambda$ is either too low or too high. 
This indicates that an optimal balance in weighting conflicts is critical---under weighting ($\lambda {\ll} 0.5$) fails to leverage disagreement signals effectively, while over weighting ($\lambda {\gg} 0.5$) can suppress correct but less frequent outputs. These findings validate our choice of $\lambda {=} 0.5$ as the default in \HYBRID. 
This trend also holds well across other datasets. 

However, when the size of the calibration dataset $N_\text{cal-base}$ varies, the peak of the curve shifts accordingly. 
For instance, when $N_\text{cal-base}$ is relatively small (=30), \CONFORMAL is significantly weaker than the conflict signal, resulting in the curve peaking at a larger $\lambda$, around 0.9. 
In contrast, when $N_\text{cal-base}$ is large (=300), the peak shifts leftward to a $\lambda$ value around 0.3-0.4.

\begin{figure}[t]
  \centering
  \captionsetup{font=small, skip=0pt}
  \centering
  \includegraphics[width=0.45\textwidth]{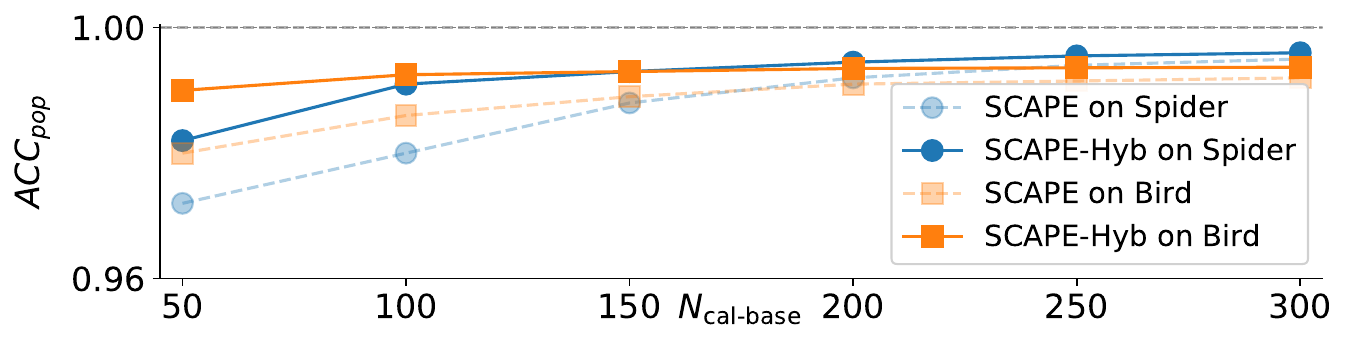}
  \caption{
    Data population accuracy $\textit{ACC}_\textit{pop}$ of \CONFORMAL and \HYBRID varying calibration dataset size $N_\text{cal-base}$, under $\textit{FPR}_\textit{pop}{=}0.2$. 
  }
  \label{fig:exp_diff_calsize}
\end{figure}

\begin{figure}[t]
  \centering
  \captionsetup{font=small, skip=0pt}
  \centering
  \includegraphics[width=0.45\textwidth]{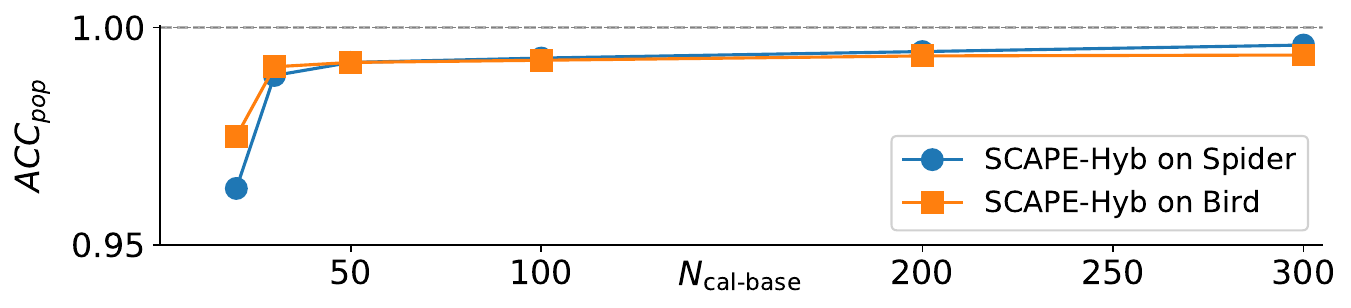}
  \caption{
    Data population accuracy $\textit{ACC}_\textit{pop}$ varying training dataset size $N_\text{cls}$, under $\textit{FPR}_\textit{pop}{=}0.2$. 
  }
  \label{fig:exp_diff_trainsize}
\end{figure}

\begin{figure}[t]
  \centering
  \captionsetup{font=small, skip=1pt}
  \centering
  \includegraphics[width=0.43\textwidth]{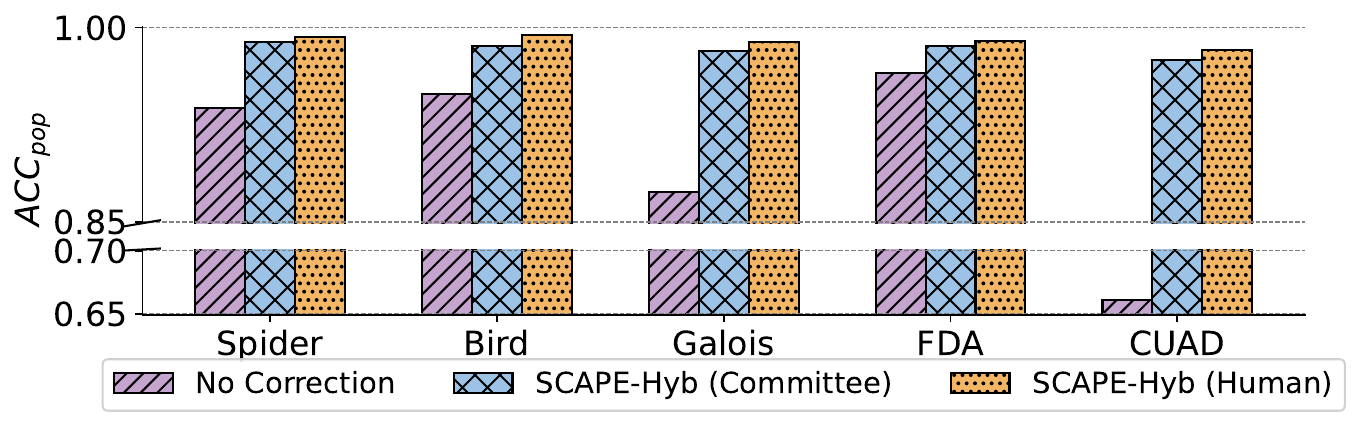}
  \caption{
    Data population accuracy $\textit{ACC}_\textit{pop}$ using human-annotated vs. LLM committee-generated training data.
  }
  \label{fig:exp_human}
\end{figure}

% \begin{table}[t]
%   \centering \small
%   \captionsetup{font=small, skip=4pt}
%   \caption{
%     Comparison of Single- and Two-Phase Schema Discovery.
%   }
%   \label{tab:exp_schema_ablation}
%   \begin{tabular}{l l c c c}
%   \toprule
%   \textbf{Dataset} & \textbf{Method} & \textbf{Invalids} & \textbf{$\textit{Rec}_\textit{sch}^\textit{attr}$} & \textbf{$\textit{Pre}_\textit{sch}^\textit{attr}$} \\
%   \midrule
%   \multirow{3}{1cm}{\SPIDER} 
%     & General-only & 2 & 0.989 & 0.522 \\
%     & Query-specific-only & 12 & 0.951 & 0.968 \\
%     & \textbf{Two-phase (Ours)} & \textbf{1} & \textbf{0.991} & \textbf{0.956} \\
%   \midrule
%   \multirow{3}{1cm}{\BIRD} 
%     & General-only & 0 & 1.000 & 0.371 \\
%     & Query-specific-only & 9 & 0.906 & 0.971 \\
%     & \textbf{Two-phase (Ours)} & \textbf{0} & \textbf{1.000} & \textbf{0.942} \\
%   \bottomrule
%   \end{tabular}
% \end{table}

\begin{table}[t]
  \centering \small
  \captionsetup{font=small, skip=4pt}
  \caption{
    Evaluation of Schema Discovery over All Datasets. 
    \textnormal{\textit{\# Invalids} refers to the number of schema discoveries missing essential attributes, such that the query cannot be answered. This is across all queries in Table \ref{tab:exp_datasets}}.
  }
  \label{tab:exp_schema_ablation}
  \begin{tabular}{p{4.1cm} c c c}
  \toprule
  \textbf{Method} & \textbf{\# Invalids} & \textbf{$\textit{Rec}_\textit{sch}^\textit{attr}$} & \textbf{$\textit{Pre}_\textit{sch}^\textit{attr}$} \\
  \midrule
    ISD (Phase I Only) & 2 & 0.989 & 0.522 \\
    ISD (Phase II Only) & 12 & 0.951 & 0.968 \\
    ISD (Phase I \& II) & 1 & 0.991 & 0.956 \\
    \textbf{ISD (Phase I \& II + Repair)} & \textbf{0} & \textbf{1.000} & \textbf{0.956} \\
  \bottomrule
  \end{tabular}
\end{table}

\subsubsection{Effect of Calibration Dataset Size $N_\text{cal-base}$.}
\label{sec:exp_population_cal_size}
~
Figure~\ref{fig:exp_diff_calsize} illustrates the impact of calibration dataset size ($N_\text{cal-base}$, defined in \S\ref{sec:scape}) on data population accuracy ($\textit{ACC}_\textit{pop}$) for both \CONFORMAL and \HYBRID (measured at $\textit{FPR}_\textit{pop}{=}0.2$). 
We report results only on \SPIDER and \BIRD for brevity. 
%as the other datasets (\Galois, \FDA, and \CUAD) do not contain enough data examples (i.e., rows or document chunks) to meaningfully vary calibration size.
Across the reported datasets, we observe a steady improvement in accuracy as $N_\text{cal-base}$ increases, with performance plateauing once the calibration set exceeds roughly 100-150 examples. 

This finding demonstrates that near-optimal accuracy can be attained with only a modest calibration set. 
For example, \HYBRID achieves accuracy above 0.99 on \SPIDER with just 150 calibration examples. 
These results demonstrate the data efficiency of our method, as high accuracy is attainable even with limited calibration data, an advantage in low-resource or high-cost settings.

Furthermore, \HYBRID consistently surpasses \CONFORMAL when calibration data is scarce. By leveraging conflict information, \HYBRID compensates for the reduced supervision in small calibration sets, reinforcing its robustness in such scenarios.

\subsubsection{Effect of Training Dataset Size $N_\text{cls}$.}
\label{sec:exp_population_train_size}
~
Figure~\ref{fig:exp_diff_trainsize} demonstrates that data population accuracy ($\textit{ACC}_\textit{pop}$) improves with larger training set size ($N_\text{cls}$), but plateaus quickly---for example, reaching ${>}0.99$ accuracy with just 50 examples. These results suggest that our approach is label-efficient: high accuracy can be achieved with limited training data and scales well as more data becomes available.

\subsubsection{Impact of Label Source: Human vs. LLM-Synthetic.}
\label{sec:exp_population_label_source}
~
Figure~\ref{fig:exp_human} compares data population accuracy ($\textit{ACC}_\textit{pop}$) when the training labels used in correction are sourced from either human annotations or LLM committee decisions. 
\HYBRID performs comparably under both label sources, with ${<}1\%$ difference in accuracy. This suggests that high-quality synthetic labels from LLM committees can be a viable alternative to human annotations in training error detection classifiers. 
Compared to data sourced from humans, LLM-generated labels are relatively cheaper, easier to obtain, and more practical for scaling to large datasets.

\subsection{Results of Schema Discovery}
\label{sec:exp_schema}

We evaluate the schema discovery stage (ISD) over all datasets. 
As shown in Table~\ref{tab:exp_schema_ablation}, the final pipeline (Phase I \& II + Repair, see \S\ref{sec:schema}) achieves perfect sufficiency (identifies all required attributes) on all queries, with zero invalid cases---i.e., every extracted schema contains all necessary attributes required to answer the query, resulting in an average recall of $\textit{Rec}_\textit{sch}^\textit{attr} {=} 1.000$. 
In addition, the extracted schemas remain concise, with an average precision of $\textit{Pre}_\textit{sch}^\textit{attr} {=} 0.956$, meaning only a few redundant attributes are occasionally included---improving the efficiency of downstream data population.

\subsubsection{Ablation Study of ISD Components.}
\label{sec:exp_schema_ablation}
~
We conduct an ablation study to assess the contribution of each ISD component. Table~\ref{tab:exp_schema_ablation} compares four configurations derived from Section \ref{sec:schema}:
\begin{itemize}[itemsep=0pt, topsep=0pt, leftmargin=10pt]
  \item 
  Phase I Only: A query-independent document-based schema extraction approach that achieves high recall (0.989)---slightly below 1.0, as some query-relevant attributes are deeply embedded in the text and not easily surfaced by a general approach. However, it suffers from very low precision (0.522), as it tends to include many irrelevant attributes not required by the query.
  \item 
  Phase II Only: A query-specific strategy that improves precision (0.968) but often misses necessary attributes due to the lack of contextual information, resulting in many invalid schemas.
  \item 
  Phase I \& II: Combining both phases significantly improves overall effectiveness, achieving high recall (0.991) and precision (0.952), and greatly reducing invalid extractions.
  \item 
  Phase I \& II + Repair: A repair step further eliminates remaining invalid cases, achieving 100\% valid extractions.
\end{itemize}
These results highlight the effectiveness of the two-phase design in achieving both schema completeness and compactness, while the repair step ensures full robustness across diverse query scenarios.

% \begin{figure}[t]
%   \centering
%   \captionsetup{font=small, skip=1pt}
%   \centering
%   \includegraphics[width=0.49\textwidth]{fig/density2acc.pdf}
%   \caption{
%     Data population accuracy ($\textit{ACC}_\textit{pop}$) of \HYBRID under different information density levels, under $\textit{FPR}_\textit{pop}=0.2$. 
%   }
%   \label{fig:exp_diff_density}
%   \vspace{-10pt}
% \end{figure}

\begin{figure}[t]
  \centering
  \captionsetup{font=small, skip=0pt}
  \begin{minipage}[t]{0.50\columnwidth}
    \centering
    \includegraphics[width=\linewidth]{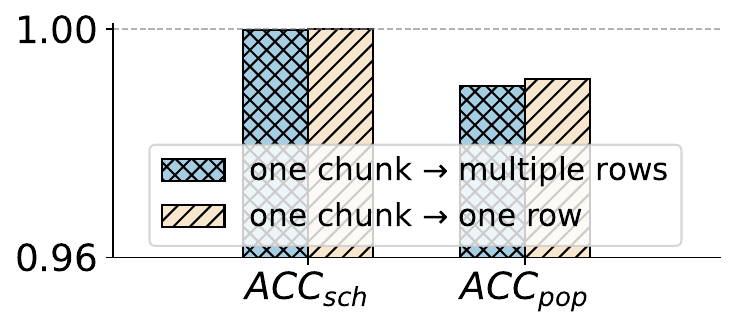}
    \caption{Schema discovery accuracy $\textit{ACC}_\textit{sch}$ and data population accuracy $\textit{ACC}_\textit{pop}$ under one-to-many chunk-to-table setting.} 
    \label{fig:exp_one_to_many}
  \end{minipage}
  \hfill
  \begin{minipage}[t]{0.45\columnwidth}
    \centering
    \includegraphics[width=\linewidth]{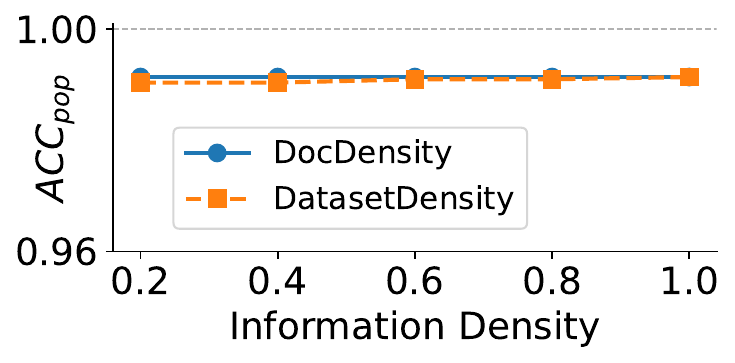}
    \caption{
      Data population accuracy ($\textit{ACC}_\textit{pop}$) of \HYBRID under diff information density levels, under $\textit{FPR}_\textit{pop}$=0.2. 
    }
    \label{fig:exp_diff_density}
  \end{minipage}
\end{figure}

\subsection{Additional Analyses}
\label{sec:exp_additional}

\subsubsection{Results under One-to-Many Chunk-to-Table Mapping.}
\label{sec:exp_population_onetomany}
~
We created a modified \SPIDER dataset, merging two or three document chunks with different schema to simulate one-to-many mappings, where each chunk contributes to multiple schemas, contrasting with the default one-to-one setup. As shown in Figure~\ref{fig:exp_one_to_many}, schema discovery accuracy ($\textit{ACC}_\textit{sch}$) remains unchanged, while data population accuracy ($\textit{ACC}_\textit{pop}$) drops slightly by 0.13\% due to increased ambiguity when mapping chunks to multiple rows. 
This trend holds across datasets. Since our system processes attributes iteratively, performance stays robust if schema descriptions are clear, emphasizing the need for precise schema discovery. The results confirm our method's ability to handle one-to-many mappings when schemas are well-defined.

\subsubsection{Impact of Information Density.}
\label{sec:exp_population_info_density}
~
Figure~\ref{fig:exp_diff_density} tests \MAINNAME's performance under different information density levels, examining two aspects: DocDensity, which quantifies how much of a document chunk's text is relevant to the extracted data, and DatasetDensity which asseses how many document chunks in the dataset are query-relevant (both presented as fractions with 1 being the most relevant and lower values signifying increased presence
of irrelevant information aiming to deter correct inference).
\MAINNAME maintains high accuracy across all scenarios.
Our key findings are that DocDensity has minimal effect because \MAINNAME processes attributes independently during TDP, avoiding confusion from irrelevant content. In addition DatasetDensity has little impact as \MAINNAME's ISD stage performs global schema discovery, easily filtering irrelevant chunks.
These results show \MAINNAME works well with real-world data where relevant information may be sparse or unevenly distributed.

\subsubsection{Token and Performance Considerations.}
\label{sec:exp_token_consumption}
~
This study demonstrates that near-perfect accuracy for unstructured data query processing is achievable. While monetary cost modeling and query performance remain critical considerations, they are beyond the scope of this work. They do represent key directions for future research however. 

In all cases in our experiments total per query execution for \HYBRID (including schema discovery, data extraction, automated training and calibration data acquisition, classifier training, calibration and query execution) is under 4 hours. This time is roughly broken down as follows: approximately 40 minutes for schema discovery on average, which is not optimized currently and sequentially processes all documents two times (via GPT-5 API \cite{openai_gpt5}); on average 3 hours per query to extract the data using a multi-threaded setting to parallelize the process without any inference optimizations (on a 8 A100 GPU cluster) and around 20 minutes for data correction, which is a manual process in our implementation currently. Token consumption is not currently optimized. For the \Galois dataset, average token consumption per query execution is approximately 18M.

There is ample scope for optimizations. This could involve deploying techniques from the literature, such as sampling to reduce the number of documents processed for schema discovery, leveraging LLMs to automatically generate code for efficient data extraction, thereby minimizing token usage as well as using advanced LLMs for correction. 
In all cases, however, formally quantifying the trade-offs between these optimizations and accuracy remains a primary focus.

\section{Related Work}
\label{sec:related}

LLMs have enabled novel lines of research to realize the vision of query processing on unstructured data. \cite{arora2023language} presents EVAPORATE, a system that uses Large Language Models (LLMs) to automatically generate structured, queryable tables from heterogeneous semi-structured documents. It explores direct extraction and a more cost-effective code synthesis approach (EVAPORATE-CODE+) that generates and aggregates multiple code snippets using weak supervision to achieve higher quality. TWIX is a tool that extracts structured data from templatized documents by first inferring the underlying visual template \cite{lin2025twix}. ZENDB, is a document analytics system designed to answer ad-hoc SQL queries on collections of templatized documents \cite{lin2024accurateefficientdocumentanalytics}.

\cite{patel2025semanticoperatorsdeclarativemodel} introduces semantic operators, a novel declarative model for AI-based data transformations that extend traditional relational operators with natural language specifications. Dai et. al., \cite{dai2024uqequeryengineunstructured} 
proposed UQE (Universal Query Engine), a new system designed for flexible and efficient analytics on unstructured data using a dialect of SQL called UQL (Universal Query Language). Liu et. al., \cite{liu2025palimpchat, liu2025palimpzest} introduce a declarative system for optimizing AI workloads, particularly "Semantic Analytics Applications" (SAPPs), which interleave traditional data processing with AI-driven semantic reasoning. \cite{urban2024efficientlearnedqueryexecution} introduces ELEET, a novel execution engine that enables seamless querying and processing of text as a first-class citizen alongside tables by leveraging learned multi-modal operators (MMOps). \cite{biswal2024text2sqlenoughunifyingai} proposes Table-Augmented Generation (TAG) as a unified paradigm for answering natural language questions over databases, addressing the limitations of existing Text2SQL and RAG methods by integrating language model reasoning with database capabilities. Anderson et. al., \cite{anderson2024designllmpoweredunstructuredanalytics} introduce infrastructure for unstructured data analytics. Wang \cite{wang} presents Unify, an innovative system that leverages large language models (LLMs) to automatically generate, optimize, and execute query plans for unstructured data analytics queries expressed in natural language. \cite{10.1145/3626246.3654732} proposes CAESURA, a novel query planner that utilizes large language models (LLMs) to translate natural language queries into executable multi-modal query plans, which can include complex operators for various data modalities beyond traditional relational data. \cite{DBLP:journals/debu/00020F25} is a novel LLM-powered analytics system designed to handle data analytics queries over multi-modal data lakes by taking natural language queries as input, orchestrating a pipeline, and outputting results. Recently \cite{satriani2025logical}, presented Galois a system for logical and physical optimization for SQL execution over LLMs and demonstrated optimization strategies to reduce token costs with small accuracy implications compared to other approaches. Our work is the first to propose a formal framework to quantify query accuracy execution over unstructured data with guarantees.
\section{Conclusion and Future Work}
\label{sec:con}

This paper introduces \MAINNAME, a novel framework for executing error-aware analytical queries over unstructured data. \MAINNAME features a two-stage pipeline that first discovers a query-specific schema and then populates relational tables.  A key contribution is the introduction of \CONFORMAL and \HYBRID, statistically calibrated methods for error detection that provide coverage guarantees.  Experimental results demonstrate that \MAINNAME significantly reduces data extraction errors from as high as 30\% to less than 1\%, while ensuring high schema completeness with 100\% recall. 
The core reliability framework presented raises numerous promising avenues for future work, ranging from incorporating \MAINNAME to other unstructured data query processing frameworks \cite{10.1145/3626246.3654732}, to designing novel query processing techniques to enhance performance, optimizing for monetary cost, and mitigating bounded errors in extraction results.

\bibliographystyle{ACM-Reference-Format}
\bibliography{base}

\tech{\appendix

\section{Appendix: Prompt Templates Used in \MAINNAME}

We include the prompt templates used in different components of \MAINNAME for reproducibility and future work reference.

\subsection{General Schema Discovery Prompt}
\label{apx:prompt_general}

This prompt is used to iteratively construct a query-agnostic schema (ISD Phase I, \S\ref{sec:schema}) over the document chunks. 
The actual prompt includes a one-shot example of input and output to guide the model. For brevity, the example is omitted here.

\begin{myquote}
You are a database expert specializing in relational schema design for heterogeneous, natural-language documents. Given a document and the current schema state, your goal is to iteratively construct and refine a schema that accurately captures the document's structure and semantics.

\textbf{Instructions:}
\begin{itemize}[itemsep=0pt, topsep=0pt, leftmargin=0pt]
  \item Identify all salient attributes from the document.
  \item Determine whether the current schema can accommodate the document; create a new table if more than two attributes are missing from any suitable table.
  \item Assign the document to the most appropriate table in the schema.
  \item For each attribute, provide a 1--2 sentence context explanation derived from its usage in the document.
\end{itemize}

\textbf{Rules:}
\begin{itemize}[itemsep=0pt, topsep=0pt, leftmargin=0pt]
  \item Only include attributes that can be supported by explicit evidence in the document.
  \item Avoid adding placeholders like \texttt{ID} unless directly mentioned or inferable.
  \item Provide clear step-by-step reasoning before outputting the revised schema and document assignment.
\end{itemize}

\textbf{Format:}
\begin{itemize}[itemsep=0pt, topsep=0pt, leftmargin=0pt]
  \item \textbf{Input:} \{ "Document": <text>, "Record of Schema": <schema state> \}
  \item \textbf{Output:} \{ "Reasoning": <text>, "Updated Record of Schema": <tables>, "Assignment": <table name> \}
\end{itemize}
\end{myquote}

\subsection{Query-Specific Schema Discovery Prompt}
\label{apx:prompt_query}

This prompt is used to iteratively construct a query-specific schema (ISD Phase II, \S\ref{sec:schema}) tailored for the given query. 
The actual prompt includes a one-shot example to guide behavior. The example is omitted here for brevity.

\begin{myquote}
You are a database expert focusing on schema design. Your task is to iteratively construct a query-specific relational schema over a collection of natural-language documents. In each iteration, you are given:
\begin{itemize}[itemsep=0pt, topsep=0pt, leftmargin=0pt]
  \item a natural language query,
  \item a new document,
  \item a general schema derived from the entire corpus, and
  \item the current query-specific schema state.
\end{itemize}

\textbf{Instructions:}
\begin{itemize}[itemsep=0pt, topsep=0pt, leftmargin=0pt]
  \item Determine which table (from the general schema) the document maps to.
  \item Identify any query-relevant attributes from the document that are necessary to answer the query, including those needed for joins.
  \item Update the query-specific schema only by:
    \begin{itemize}[itemsep=0pt, topsep=0pt, leftmargin=15pt]
      \item[(a)] adding a new table (reusing the general schema name), or
      \item[(b)] adding new attributes to an existing table.
    \end{itemize}
  \item You may apply \emph{at most one} of these actions per iteration.
  \item Assign the document to a table, or return \texttt{"Assignment": None} if the document is irrelevant.
\end{itemize}

\textbf{Rules:}
\begin{itemize}[itemsep=0pt, topsep=0pt, leftmargin=0pt]
  \item Do not assess whether the data satisfies the query conditions—only whether it contains schema-relevant attributes.
  \item Do not include unnecessary attributes.
  \item For aggregate queries, include raw attributes necessary for computing the aggregate.
\end{itemize}

\textbf{Format:}
\begin{itemize}[itemsep=0pt, topsep=0pt, leftmargin=0pt]
  \item \textbf{Input:} \{ "Document", "Query", "Record of Query-specific Schema", "General Schema" \}
  \item \textbf{Output:} \{ "Reasoning", "Updated Record of Query-specific Schema", "Assignment" \}
\end{itemize}

\end{myquote}

\subsection{Tabular Data Population Prompt (TDP)}
\label{apx:prompt_datapop}

Two prompts are used: one to select the target table for a document chunk, and another to extract values for each attribute.

\paragraph*{(1) Table Resolver Prompt}

\begin{myquote}
You are a database expert. Your task is to determine which table a given document belongs to, based on a provided set of table schemas. Each document can be assigned to only one table.

\textbf{Instructions:}
\begin{itemize}[itemsep=0pt, topsep=0pt, leftmargin=0pt]
  \item Read the document and compare it with the attribute descriptions in each table schema.
  \item Assign the document to the table whose schema best matches its content.
\end{itemize}

\textbf{Format:}
\begin{itemize}[itemsep=0pt, topsep=0pt, leftmargin=0pt]
  \item \textbf{Input:} \{ "Document", "Schema" \}
  \item \textbf{Output:} \{ "Table Assignment": <Schema Name> \}
\end{itemize}
\end{myquote}

\paragraph*{(2) Attribute Extractor Prompt}

\begin{myquote}
You are a database expert. Your task is to extract a specific attribute value from a natural language document, given a table schema and a target attribute.

\textbf{Instructions:}
\begin{itemize}[itemsep=0pt, topsep=0pt, leftmargin=0pt]
  \item Examine the document and the schema.
  \item Locate the value in the document corresponding to the target attribute.
  \item If the attribute value is found, return it; otherwise, return \texttt{None}.
\end{itemize}

\textbf{Format:}
\begin{itemize}[itemsep=0pt, topsep=0pt, leftmargin=0pt]
  \item \textbf{Input:} \{ "Document", "Schema", "Target Attribute" \}
  \item \textbf{Output:} \{ <Target Attribute>: <Extracted Value or None> \}
\end{itemize}
\end{myquote}

\subsection{Dataset Document Generation Prompt}
\label{apx:prompt_dataset}

In ReDD-S and ReDD-B (\S\ref{sec:exp}), document chunks were generated by transforming structured data (e.g., table rows) into natural language using the following prompt template.

\begin{myquote}
You are a technical writer tasked with converting structured table rows into natural, paragraph-style descriptions suitable for a professional document corpus (e.g., medical reports, financial filings, legal records).

\textbf{Instructions:}
\begin{itemize}
  \item Given a structured data row and a schema, rewrite the row as a coherent, grammatically correct paragraph.
  \item Preserve factual accuracy and semantics.
  \item Implicitly include all attributes without listing them as key–value pairs.
  \item The generated text should resemble real-world document language, using varied sentence structure and terminology.
\end{itemize}

\textbf{Format:}
\begin{itemize}
  \item \textbf{Input:} \{ "Schema": <table schema>, "Row": <structured data> \}
  \item \textbf{Output:} Natural language paragraph describing the row.
\end{itemize}
\end{myquote}

}

\end{document}